\documentclass[12pt]{article}
\usepackage{amssymb,diagrams}
\newcommand{\be}{\begin{eqnarray}}
\newcommand{\ee}{\end{eqnarray}}
\newcommand{\pa}{\partial}
\newcommand{\vt}{\vartheta}
\renewcommand{\d}{{\rm d}}
\newcommand{\dl}{\delta}
\newcommand{\td}{\tilde{\d}}
\newcommand{\D}{{\rm D}}
\newcommand{\tdl}{\tilde{\delta}}
\newcommand{\tD}{\tilde{\rm D}}
\newcommand{\Dl}{{\cal D}}

\newcommand{\vari}{{\delta \!\!\!{}^-}}

\title{\bf Bicomplexes and B\"acklund transformations}
\date{  }
\author{A. Dimakis$^1$ and F. M\"uller-Hoissen$^2$}

\begin{document}
\renewcommand{\theequation} {\arabic{section}.\arabic{equation}}
\maketitle

\begin{center}
 $^1$ Department of Financial and Management Engineering, 
 University of the Aegean, \\ 
 31 Fostini Str.,
 GR-82100 Chios, dimakis@aegean.gr
\vskip.1cm
 $^2$ Max-Planck-Institut f\"ur Str\"omungsforschung,
 Bunsenstrasse 10, D-37073 G\"ottingen, 
 fmuelle@gwdg.de
\end{center}

\begin{abstract}
A bicomplex is a simple mathematical structure, in particular associated 
with completely integrable models. The conditions defining a bicomplex 
are a special form of a parameter-dependent zero curvature condition. 
We generalize the concept of a Darboux 
matrix to bicomplexes and use it to derive B\"acklund transformations 
for several models. The method also works for Moyal-deformed 
equations with a corresponding deformed bicomplex.
\end{abstract}

\section{Introduction}
\setcounter{equation}{0}
B\"acklund transformations arose in the 19th century in a differential 
geometric context \cite{Eisen09}. Despite of the fact that this is already a 
rather old subject, it is still the subject of quite a lot of recent articles. 
Indeed, in the case of many nonlinear equations, a B\"acklund transformation 
(BT) remains the only hope to construct sufficiently complicated exact solutions. 
\vskip.1cm

The essence of the concept of a BT is basically the following 
(see \cite{BT,Roge+Shad82,Herm,Draz+John89}, for example). 
Suppose we have two (systems of) partial differential 
equations\footnote{We may consider difference equations as well.}
$EQ_1[u_1]=0$ and $EQ_2[u_2]=0$, depending on a variable $u_1$ and 
its partial derivatives, respectively $u_2$ and its partial derivatives. 
A BT is then given by relations between the two 
variables and their partial derivatives which determine $u_2$ in terms of $u_1$ 
such that $EQ_2[u_2]=0$ if $EQ_1[u_1]=0$ holds. So, given a solution 
of $EQ_1=0$, it determines a corresponding solution of $EQ_2=0$. This will 
only be of help, of course, if the relations between $u_1$ and $u_2$ are 
considerably simpler than the equations $EQ_1=0$ and $EQ_2=0$, and if 
some solutions of one of these equations are known. 
 For instance, $EQ_1=0$ and $EQ_2=0$ 
may be higher order partial differential equations and the BT 
only of first order. 
If $EQ_1=EQ_2$, then such a transformation is called an {\em auto-B\"acklund 
transformation}. It can be used to generate new solutions of $EQ_1=0$ 
from given ones. How to find (useful) BTs? 
 For most completely integrable models, ways to construct BTs are known. 
The existence of such a transformation is usually taken as a criterion 
for complete integrability. 
\vskip.1cm

A large class of completely integrable equations in two (space-time) 
dimensions admits a {\em zero curvature formulation} 
\be
   U_t -V_x + [U,V] = 0    \label{zero_curv}
\ee
(see  \cite{Fadd+Takh87}, in particular) with matrices $U(u, \lambda)$ and 
$V(u, \lambda)$ depending on a parameter $\lambda$, besides the dependent 
variable $u$. Here, $U_t$ and $V_x$ denote the partial derivatives 
of $U$ and $V$ with respect to the coordinates $t$ and $x$, respectively. 
(\ref{zero_curv}) is the compatibility condition of the linear system
\be
   z_x = U(u,\lambda) \, z \, , \qquad z_t = V(u, \lambda) \, z 
                        \label{linear_system}
\ee  
with a (matrix) `wave function' $z$. 
 A special class of auto-BTs is then 
obtained via $\lambda$-dependent transformations
\be
   z' = Q(u,u',\lambda) \, z
\ee
depending on two fields, $u$ and $u'$, which preserve the form of 
the linear system, so that
\be
   z'_x = U(u',\lambda) \, z' \, , \qquad z'_t = V(u', \lambda) \, z' \; .
                 \label{linear_system'}
\ee
Then, if $u$ is a solution of the equation modelled by the zero curvature 
condition, $u'$ is also a solution. $Q$ has to satisfy the equations 
\be
   Q_x = U(u') \, Q - Q \, U(u) \, , \qquad
   Q_t = V(u') \, Q - Q \, V(u)   \label{UV_gt}
\ee
and is called {\em Darboux matrix} 
\cite{Cies98,Levi86}.\footnote{This is sometimes called a {\em Darboux 
transformation} (see \cite{Ford94}, for example), but should not 
be confused with the classical Darboux transformation 
\cite{Matv+Sall91} which, however, is indeed related to a 
Darboux matrix \cite{Levi86}.} 
It is not known, except for special examples, whether all auto-BTs 
of an integrable model, possessing a zero curvature formulation, 
can be recovered in this way. Many auto-BTs 
are known to be of this Darboux form, however (see 
\cite{Cies98,BT-gauge}, for example). 
In many cases, an ansatz for $Q$ suffices which is linear in $\lambda$ 
\cite{Levi86}. 
The important `dressing method' of Zakharov and Shabat 
\cite{Zakh+Shab80} actually involves the construction of a Darboux matrix.
\vskip.1cm

In terms of the $\lambda$-dependent `covariant derivative' 
$D = d - U \, dx - V \, dt = d + A$, 
where $d$ is the exterior derivative on $\mathbb{R}^2$ (with coordinates 
$x$ and $t$), (\ref{linear_system}) and (\ref{linear_system'}) can be 
written as $D z = 0$ and $D'z' = 0$, respectively, and 
(\ref{zero_curv}) becomes $D^2 = 0$ which is 
$F = d A + A \wedge A = 0$.\footnote{Without further conditions, $F=0$ 
is simply solved by $A = g^{-1} \, dg$ with an invertible matrix $g$. 
If we require the dependence of $A$ on $\lambda$ to be polynomial or 
rational, this results in nontrivial equations, however.}  
(\ref{UV_gt}) can be rewritten as
\be
     A' \, Q = Q \, A - d Q 
\ee
which, for an invertible $Q$, is the transformation law of a connection 
(`gauge potential') under a gauge transformation given by $Q$. This is 
equivalent to the covariance property
\be
     D' \, Q = Q \, D  \; .
\ee
of the covariant derivative. This scheme can be generalized to hetero-BTs 
where $Q$ relates two different zero curvature equations.\footnote{Various 
relations between zero curvature conditions and BTs have also 
been discussed in \cite{BT,Roge+Shad82,Herm,BT-0curv}, for example.} 
\vskip.1cm

In the special case where $D$ depends linearly on $\lambda$, it naturally 
splits into two linear operators which do not depend on $\lambda$ and 
we have an example of a bicomplex \cite{DMH-bico} (and even a 
bi-differential calculus \cite{DMH-bidiff}). 
A {\em bicomplex} is an ${\mathbb{N}}_0$-graded linear 
space (over ${\mathbb{R}}$) $M = \bigoplus_{s \geq 0} M^s$ 
together with two linear maps $\Dl , \D \, : \, M^s \rightarrow M^{s+1}$ 
satisfying
\be
   \Dl^2 = 0 \, , \qquad  \D^2 = 0 \, , \qquad  
   \D \, \Dl + \Dl \, \D = 0  \; .    \label{bicomplex_cond}
\ee 
Introducing an auxiliary real parameter $\lambda$, these equations 
can be written as a generalized parameter-dependent zero curvature 
condition,\footnote{There are examples in section 4 where 
the bicomplex maps $\Dl$ and $\D$ involve complex quantities but are 
{\em not} linear over ${\mathbb{C}}$. The bicomplex equations can 
then only be cast into this form for {\em real} $\lambda$.}
\be
    (\Dl - \lambda \, \D)^2 = 0  \; .   
\ee
Typically, the maps $\Dl$ and $\D$ depend on a (set of) variable(s) $u$ 
such that the bicomplex conditions hold if and only if $u$ is a solution of a 
system of (e.g., partial differential) equations. 
The corresponding linear system is 
\be
    (\Dl - \lambda \, \D) \, z = 0   
\ee 
(for $z \in M^0$), or a slight modification of it. The simple linear 
$\lambda$-dependence of this linear system has to be contrasted 
with the, in general, nonlinear $\lambda$-dependence of $U$ and $V$ in 
(\ref{linear_system}). Thus, at first sight, the bicomplex 
formulation looks like a severely restricted zero curvature condition.\footnote{Actually  
many integrable models can be derived by reduction of the self-dual Yang-Mills 
equation which possesses a zero curvature formulation linear in $\lambda$.}
However, rewriting the bicomplex equations, if possible, 
in the form (\ref{zero_curv}) and (\ref{linear_system}), in general results 
in a nonlinear $\lambda$-dependence of the corresponding $U$ and $V$.  
But in general it will not be possible to rewrite a given zero curvature formulation, 
showing a nonlinear $\lambda$-dependence, in bicomplex form\footnote{Clearly, 
this applies in particular to zero curvature equations with a non-rational 
dependence on $\lambda$ (see \cite{Fadd+Takh87}, 
for example).}, although it is possible that a bicomplex formulation exists for this 
model. Disregarding the $\lambda$-dependence, the structure of the bicomplex 
equations is somewhat less restrictive than (\ref{zero_curv}). 
In a series of papers 
\cite{DMH-bico,DMH-bidiff,DMH-KdV,DMH-ncKdV,DMH-ncbico,DMH-FK,Gueu01}, 
several known integrable models have been cast into the bicomplex form, 
some new models have been constructed, and it has been demonstrated, 
in particular, how conservation laws can be derived from it in the case of 
evolution-type equations. 
A bridge from bi-Hamiltonian systems to bicomplexes has 
been established in \cite{bico-biHam}. It should be stressed, however, that 
a bicomplex structure is much more general and does not presuppose the 
existence of a symplectic or Hamiltonian structure.
\vskip.1cm

The idea of a Darboux matrix is immediately carried over to bicomplexes. 
Let ${\cal B}_i = (M_i, \Dl_i, \D_i)$, $i=1,2$, be two bicomplexes 
depending on variables $u_1$ and $u_2$, respectively. 
We propose the following definition. 
\vskip.1cm
\noindent
{\em Definition.} A {\em Darboux-B\"acklund transformation} (DBT) for the two 
bicomplexes ${\cal B}_1$ and ${\cal B}_2$ is given by a $\lambda$-dependent 
linear operator $Q(\lambda) : M_1 \rightarrow M_2$ such that\footnote{The 
requirement of `form-invariance' of $U$ and $V$ in the Darboux matrix 
formalism is here replaced by the requirement that $Q$ preserves the 
linear $\lambda$-dependence of the operator $\Dl - \lambda \, \D$.} 
\be
   (\Dl_2 - \lambda \, \D_2) \, Q(\lambda) 
 = Q(\lambda) \, (\Dl_1 - \lambda \, \D_1)  \label{DBT}
\ee
for all $\lambda$. This is an {\em auto-Darboux-B\"acklund transformation} 
if the two bicomplexes are associated with the same equation.\footnote{If 
$Q$ does not depend on $\lambda$ and is invertible, an auto-DBT reduces 
to an equivalence transformation of the two bicomplexes.} 
\hfill                $\diamondsuit$
\vskip.1cm

Since the bicomplex maps depend on the fields $u_1$ and $u_2$ (which are 
solutions of the respective field equations), so does $Q$ as a consequence of 
(\ref{DBT}). It implies
\be
      Q(\lambda) \, (\Dl_1 - \lambda \, \D_1)^2 
   = (\Dl_2 - \lambda \, \D_2)^2 \, Q(\lambda)    \; .
\ee
Hence, if $u_1$ is a solution of the equation associated with ${\cal B}_1$, then 
$(\Dl_2 - \lambda \, \D_2)^2 \, Q(\lambda) = 0$. If $Q(\lambda)$ is invertible, 
an assumption which we will make in all our examples,
this implies that $u_2$ is a solution of the equation associated with ${\cal B}_2$.
If $Q(\lambda)$ is not invertible, this cannot be concluded, in general. 
\vskip.1cm

Given three bicomplexes which are connected by DBTs 
$Q_{21} : M_1 \rightarrow M_2$ and $Q_{32} : M_2 \rightarrow M_3$, 
the composition $Q_{32} \, Q_{21}$ is also a DBT:
\be
      (\Dl_3 - \lambda \, \D_3) \, Q_{32}(\lambda) \, Q_{21}(\lambda)
  = Q_{32}(\lambda) \, Q_{21}(\lambda) \, (\Dl_1 - \lambda \, \D_1) \; .
\ee
Let us now consider four bicomplexes ${\cal B}_i$, $i=0,1,2,3$, with 
DBTs $Q_{10}, Q_{20}, Q_{31}, Q_{32}$.
Suppose we can solve the respective DBT conditions such that solutions 
$u_i$ of $EQ_i =0$ can be expressed via the DBT $Q_{ij}$ in terms of 
solutions of $EQ_j =0$. The condition
\be
      Q_{31} \, Q_{10} = Q_{32} \, Q_{20} 
                                        \label{permutability}
\ee
is then in many cases strong enough to guarantee that a solution $u_0$ of $EQ_0 =0$ 
is taken via $Q_{31} \, Q_{10}$ and also via $Q_{32} \, Q_{20}$ to the same 
solution $u_3$ of $EQ_3 =0$ (see Fig.~1). The condition 
(\ref{permutability}) is our formulation of the `permutability theorem' 
(see \cite{Eisen09,Roge+Shad82}, for example). In the case of auto-DBTs, it leads to 
nonlinear superposition rules for solutions of the respective equation. 
Typically, a BT (obtained from a DBT) depends on some arbitrary constants. The 
condition (\ref{permutability}) then enforces relations between the corresponding 
constants on the left ($u_0 \mapsto u_1 \mapsto u_3$) and the right way 
($u_0 \mapsto u_2 \mapsto u_3$) in the right diagram of Fig. 1. The usual 
formulation of a permutability theorem is that if these relations between the 
constants hold, then the two ways yield the same solution $u_3$.

\diagramstyle[PostScript=dvips]
\newarrow{Mapsto}{|}{-}{}{-}>
\begin{diagram}[notextflow]
         &                           &M_3&                           &        &         &        &         &u_3&                     &     \\
         &\ruTo^{Q_{31}}&        &\luTo^{Q_{32}}&        &          &       &\ruMapsto&      &\luMapsto&      \\
 M_1&                           &        &                           &M_2 &\quad & u_1&         &      &                      &u_2 \\   
         &\luTo^{Q_{10}}&        &\ruTo^{Q_{20}}&         &          &       &\luMapsto&      &\ruMapsto&      \\
         &                          &M_0 &                           &         &          &       &       &u_0&                       &
\end{diagram}
\begin{center}
 \parbox{14.0cm}{
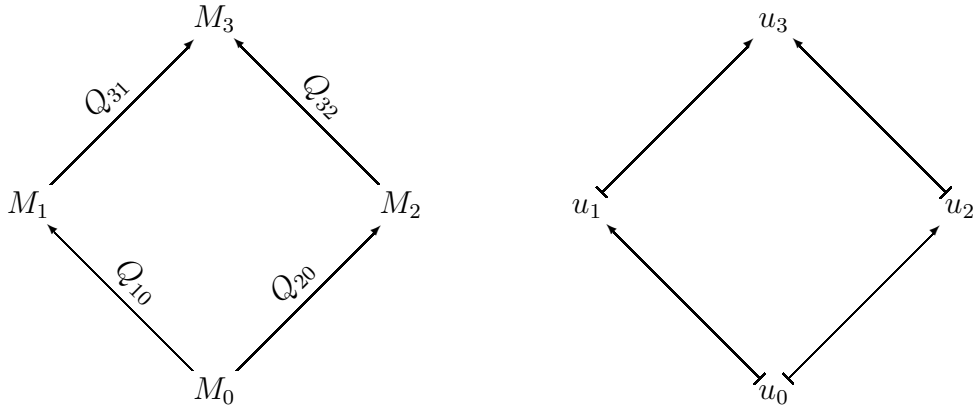
     \centerline{\bf Fig.~1:} 
     \vskip.1cm  \noindent
A diagram of DBTs and a corresponding induced mapping of solutions in case of commutativity 
of the first diagram. }
\end{center}
\vskip.1cm

The linear dependence on the parameter $\lambda$ in the bicomplex 
zero curvature formulation greatly simplifies the derivation of DBTs. 
Specializing to various models, one easily recovers well-known BTs. 
 For this purpose we make an ansatz
\be
    Q(\lambda) =  \sum_{k=0}^N \lambda^k \, Q^{(k)}  \label{Q_Nth}
\ee
with some $N \in \mathbb{N}$ and $Q^{(k)}$ not depending on $\lambda$. 
The DBT condition (\ref{DBT}) then splits into the system of equations
\be
   \Dl_2  \, Q^{(0)} -  Q^{(0)} \, \Dl_1 &=& 0  \nonumber \\
   \Dl_2  \, Q^{(k)} -  Q^{(k)} \, \Dl_1 &=&  \D_2 \, Q^{(k-1)} - Q^{(k-1)} \, \D_1 
                      \qquad (k=1, \ldots, N)    \label{NthDBT}  \\
   \D_2 \, Q^{(N)} - Q^{(N)} \, \D_1 &=& 0   \, .  \nonumber
\ee
We speak of a {\em primary DBT} when $N=1$, of a {\em secondary DBT} 
when $N=2$, and so forth.  The composition of $N$ primary DBTs is obviously 
at most an $N$-ary DBT. Generically, it will be indeed an $N$-ary DBT.\footnote{An 
exception appears in the Liouville example treated in section 2, where the composition 
of primary DBTs of the form (\ref{Liou_N=1Q}) is again a primary DBT.}
There may be higher DBTs which are not obtained in this way, however.
\vskip.1cm 

If two bicomplexes admit an {\em invertible} DBT, there is 
an equivalent DBT-problem with $Q(\lambda)$ acting on a 
single bicomplex space $M$ and $Q^{(0)} = I$, the identity 
operator.\footnote{If $Q(\lambda)$ is invertible, then 
$Q^{(0)}$ determines an equivalence transformation of bicomplexes. 
Introducing $\Dl'_2 = (Q^{(0)})^{-1} \, \Dl_2 \, Q^{(0)}$ 
and $\D'_2 = (Q^{(0)})^{-1} \, \D_2 \, Q^{(0)}$, we have 
$\tilde{Q}(\lambda) \, (\Dl_1 - \lambda \, \D_1) 
 = (\Dl'_2 - \lambda \, \D'_2) \, \tilde{Q}(\lambda) $ with 
$\tilde{Q}(\lambda) = (Q^{(0)})^{-1} \, Q(\lambda) : \, M_1 \rightarrow M_1$.}  
Expressing the dependence of $\Dl_i$ on a solution $u_i$ explicitly as 
$\Dl_i[u_i]$, the first of equations (\ref{NthDBT}) then requires 
$\Dl_1[u_1] = \Dl_2[u_2]$ for all solutions $u_1$ and $u_2$ related 
via $Q(\lambda)$. 
Let us consider the special case where the equation under consideration 
admits a bicomplex ${\cal B}_1$ such that $\Dl_1$ does not depend on $u$. 
In this case we write $\Dl_1 = \dl$ and obtain $\Dl_2 = \dl$.
The DBT-problem then takes the form $(\dl - \lambda \, \D_2[u_2]) \, Q(\lambda) 
 = Q(\lambda) \, (\dl - \lambda \, \D_1[u_1])$.
Moreover, if we look for {\em auto}-DBTs, the bicomplexes 
${\cal B}_1 = (M,\dl,\D_1)$ and ${\cal B}_2 = (M,\dl,\D_2)$ have 
to be equivalent. This means that the respective sets of bicomplex equations, 
which depend on $u$, must both be satisfied if $u$ solves the equation for 
which ${\cal B}_1$ and ${\cal B}_2$ are bicomplexes. If this holds for 
${\cal B}_1$, then an obvious way to achieve that it also holds for 
${\cal B}_2$ is to choose $\D_2[u] = \D_1[u] = \D[u]$. 
Since $\dl$ is common to both bicomplexes, we should actually hardly 
expect $\D_1[u]$ and $\D_2[u]$ to differ in a non-trivial way, although 
exceptions may exist.\footnote{If $u \mapsto \tilde{u}$ is a symmetry 
transformation of the equation for $u$, then we may also choose 
$\D_2[u] = \D_1[\tilde{u}]$. Note that a symmetry of the equation 
is not a symmetry (equivalence transformation) of an associated 
bicomplex, in general. The consequences of this observation have 
still to be explored.}
This motivates to restrict the invertible auto-DBT condition for an 
equation, which possesses a bicomplex ${\cal B} = (M,\dl,\D)$ where 
$\dl$ does not depend on $u$, to the form
\be
   (\dl - \lambda \, \D[u_2]) \, Q(\lambda) 
   = Q(\lambda) \, (\dl - \lambda \, \D[u_1])    \label{autoDBT}
\ee
with $Q^{(0)} = I$. Then we are dealing with a {\em single} 
bicomplex only. This restricted form of the auto-DBT condition 
underlies all examples in section 4. 
\vskip.1cm

How severe is the above restriction on the bicomplex ${\cal B}_1$ 
(and thus ${\cal B}$)? 
Splitting a given bicomplex map $\Dl$ as $\Dl = \dl + B$ with a 
suitable operator $\dl$ which is independent of a solution $u$ 
and satisfies $\dl^2 = 0$, the generalized curvature 
${\cal F} = [\dl , B] + B^2$ vanishes (see also section 3). 
$\cal F$ generalizes the classical differential geometric formula 
for the curvature of a connection 1-form $B$ if $\dl$ is given by 
an exterior derivative on some manifold. In that case, it is 
well-known that a gauge transformation exists which transforms 
$B$ to $B' = 0$ so that $\Dl' = \dl$. In the much more general 
bicomplex framework we do not have a corresponding theorem at hand, 
though an analogous result should be expected for relevant classes 
of bicomplexes (see section 5 for an example). This would mean that, 
if a bicomplex exists, then also a bicomplex with a map $\Dl$ which 
is independent of the solution of the underlying equation. At least, this 
motivates a corresponding ansatz. In practice, however, it is often 
difficult enough to find {\em any} bicomplex for some equation and it 
is then hardly possible to find a concrete transformation to such a special 
bicomplex. Moreover, such a transformation may change the concrete 
form of the equation (cf section 5) which then possibly makes it difficult to 
apply corresponding results (e.g., concerning DBTs) to the original 
problem.\footnote{In particular, the nonlinear Schr\"odinger equation 
(see section 4.2.7) is known to be ``gauge equivalent" to the Heisenberg 
magnet model (see \cite{Fadd+Takh87}, for example). This can be 
understood as an equivalence of bicomplexes \cite{DMH-bico,DMH-FK}.} 
 Furthermore, one should keep in mind that interesting examples 
may exist for which the above special bicomplex form cannot be reached. 
Of course, in such a case the DBT method can still be applied, although 
the calculations will be more involved, in general. 
\vskip.1cm

In \cite{DMH-ncKdV,DMH-ncbico,DMH-FK} we constructed bicomplexes for various 
Moyal-deformed classical integrable models. Here, the ordinary product of functions 
is replaced by an associative, noncommutative $\ast$-product \cite{dq}.  
The definition of a bicomplex, and in particular (\ref{bicomplex_cond}), as well 
as our definition of a bicomplex DBT still applies.
Also in this case a DBT provides us with a helpful solution generating 
technique, as will be demonstrated with an example in section 4.
\vskip.1cm

Section 2 treats the example of the Liouville equation and its discretization. 
In section 3 we elaborate our definition of a bicomplex DBT for 
a `dressed' form of the bicomplex maps \cite{DMH-bico,DMH-bidiff} and 
the restricted case where all maps act on the same bicomplex space $M$. 
Section 4 then shows how to recover auto-BTs for several well-known 
integrable models from a bicomplex DBT. Section 5 deals with the 
Harry Dym equation (see \cite{HBC89}, in particular) and  section 6 
collects some conclusions.

\section{Example: Liouville bicomplex}
\setcounter{equation}{0}
In many examples, the bicomplex space can be chosen as $M = M^0 \otimes \Lambda_n$ 
where $\Lambda_n = \bigoplus_{r=0}^n \Lambda^r$
is the exterior algebra of an $n$-dimensional real vector space with a basis $\xi^r$,
$r=1, \ldots, n$, of $\Lambda^1$. It is then sufficient
to define the bicomplex maps $\Dl$ and $\D$ on $M^0$ since via 
\be
 \D (\sum_{i_1,\ldots,i_r =1}^n \phi_{i_1 \ldots i_r} \, \xi^{i_1} \cdots \xi^{i_r} ) 
 = \sum_{i_1,\ldots,i_r =1}^n ( \D \phi_{i_1 \ldots i_r} ) \, \xi^{i_1} \cdots \xi^{i_r}
\ee
(and correspondingly for $\Dl$) they extend as linear maps to the whole of 
$M$.\footnote{In some examples, the $\xi^r$ can be realized as differentials of 
coordinates on a manifold. In this way contact is made with ordinary zero curvature 
formulations (linear in $\lambda$) of continuous integrable models. The generalization 
from the algebra of differential forms to an abstract Grassmann algebra is important, 
however, in order to treat relevant examples within the bicomplex framework.}
In the case of $\Lambda_2$, we denote the two basis elements of 
$\Lambda^1$ as $\tau, \xi$. 
They satisfy $\xi^2 = 0 = \tau^2$ and $\xi \, \tau = - \tau \, \xi$. 

\bigskip
\noindent
{\em 1. Liouville equation.} 
Let $M = C^\infty(\mathbb{R}^2, \mathbb{R}^2) \otimes \Lambda_2$.  
Let $x,y$ be coordinates on $\mathbb{R}^2$ and $z_x, z_y$ the corresponding 
partial derivatives of $z \in M^0$. We define
\be
    \Dl z = z_x \, \xi + (\sigma_+ -I) z \, \tau  \, , \qquad
  \D z  = \kappa \, e^{2 \phi} \, \sigma_- z \, \xi 
            + (z_y + \phi_y \, \sigma_3 z) \, \tau
\ee
with a constant $\kappa$, the $2 \times 2$ unit matrix $I$, and 
\be
 \sigma_3 = \left( \begin{array}{cc} 1 & 0 \\ 0 & -1 \end{array} \right) \, , \quad
 \sigma_+ = \left( \begin{array}{cc} 0 & 1 \\ 0 & 0 \end{array} \right) \, , \quad
 \sigma_- = \left( \begin{array}{cc} 0 & 0 \\ 1 & 0 \end{array} \right) \, .
\ee
Then $(M, \Dl, \D)$ is a bicomplex for the Liouville equation
\be
    \phi_{xy} = \kappa \, e^{2 \phi}   \, .  \label{Liou_eq}
\ee

Let us now consider two such bicomplexes corresponding to two different choices 
$\kappa_i$, $i=1,2$, for $\kappa$ and corresponding solutions $\phi_i$. 
Then the DBT condition (\ref{DBT}) reads 
$[\Dl , Q] = \lambda \, (\D_2 \, Q - Q \, \D_1)$.
With the ansatz (\ref{Q_Nth}) for $Q$, this becomes
\be
   [\Dl , Q^{(0)}] = 0 \, , \; 
   [\Dl , Q^{(k)}] = \D_2 \, Q^{(k-1)} - Q^{(k-1)} \, \D_1 \; (k=1, \ldots, N), \;   
   \D_2 \, Q^{(N)} = Q^{(N)} \, \D_1  \, .    \label{Liou_NthDBT}
\ee
Taking for $Q^{(k)}$ general $2 \times 2$ matrices, the entries of 
which are functions of $x$ and $y$, the first equation leads to
\be
   Q^{(0)} = a(y) \, I + b(y) \, \sigma_+   
\ee
with functions $a,b$ which do not depend on $x$.
The $k=1$ DBT condition in particular requires $a_y =0$, so that 
$a$ must be a constant. Furthermore, it leads to 
\be
     Q^{(1)} 
 = f \, I - {1 \over 2} [ b_y + b \, (\phi_1 + \phi_2)_y] \, \sigma_3 
    + c(y) \, \sigma_+ + r \, \sigma_-                    \label{Liou_Q1}
\ee
with an arbitrary function $c(y)$ and functions $f, r$ subject to
\be
  f_x &=& {b \over 2} \, ( \kappa_2 \, e^{ 2 \phi_2}- \kappa_1 \, e^{2 \phi_1} )  \\
     r &=& a \, (\phi_2 - \phi_1)_y  \, , \qquad  r_x 
          = a \, [ \kappa_2 \, e^{2 \phi_2} - \kappa_1 \, e^{2 \phi_1} ]  \; .   
                                                \label{Liou_r}   
\ee 
The next equation in (\ref{Liou_NthDBT}) requires in particular\footnote{For 
$N=1$ this is a consequence of the last equation in (\ref{Liou_NthDBT}). 
 For $N>1$, it follows from the $k=2$ equation.}  
$r_y = r \, (\phi_1 + \phi_2)_y$ and thus
\be
  r = \alpha \, e^{\phi_1 + \phi_2}
\ee
where $\alpha$ does not depend on $y$. If $a \neq 0$, the two equations 
(\ref{Liou_r}) now reproduce a well-known BT for the Liouville equation: 
\be
  (\phi_2 - \phi_1)_y = \alpha \, e^{\phi_1 + \phi_2}  \, ,  \qquad
  (\phi_1 + \phi_2)_x = {1 \over \alpha} \, (\kappa_2 \, e^{\phi_2 - \phi_1}
        - \kappa_1 \, e^{\phi_1 - \phi_2} - \alpha_x ) 
		     \label{Liou_BT}
\ee
where $a$ has been absorbed via a rescaling of $\alpha$ (cf \cite{BT}, for example).  
This is precisely obtained as the primary DBT condition with
\be
     Q = I + \lambda \, \alpha \, e^{\phi_1 + \phi_2} \, \sigma_-  \; .
                     \label{Liou_N=1Q}
\ee
The remaining freedom in the general solution for $Q$ (where $N \geq 1$) 
can only restrict this BT. In particular, it cannot lead to different BTs.
The case $a =0$ and $b \neq 0$ also does not lead to additional BTs. 

If $\kappa_1 = 1$ and $\kappa_2 = 0$, (\ref{Liou_BT}) is a hetero-BT 
connecting solutions of the Liouville equation $\phi_{xy} = e^{2 \phi}$ 
with solutions of $\phi_{xy} = 0$ (which is the wave equation in light cone 
coordinates). 

Let us now explore the permutability conditions for $N=1$. Using 
$\sigma_-^2 = 0$, (\ref{permutability}) with (\ref{Liou_N=1Q}) becomes
\be
   \phi_3 
 = \phi_0 + \ln \left( {\alpha_{10} \, e^{\phi_1} - \alpha_{20} \, e^{\phi_2} 
   \over \alpha_{32} \, e^{\phi_2} - \alpha_{31} \, e^{\phi_1}} 
                \right) \, .
\ee
This allows to compute a solution $\phi_3$ in a purely algebraic way from 
solutions $\phi_0, \phi_1,\phi_2$, if the pairs $(\phi_0, \phi_1)$ and 
$(\phi_0,\phi_2)$ satisfy (\ref{Liou_BT}).
\vskip.1cm

The fact that $N>1$ does not lead to other BTs is a special feature 
of the Liouville example. Let us consider the case $\kappa_i =1$, $i=1,2$, 
in more detail. In general, the higher DBTs should be 
expected to be compositions of primary DBTs (see also the KdV 
example in section 4.1). Indeed, in the following we show that 
the composition of two Liouville BTs is again of the form 
(\ref{Liou_BT}). Differentiating the Liouville equation (\ref{Liou_eq}) 
with respect to $x$, we find $\phi_{xxy} = 2 \phi_x \phi_{xy}$ 
and an integration with respect to $y$ leads to 
\be
 \phi_{xx} = f(x) + \phi_x{}^2 \label{Liou_phi_xx}
\ee
with integration `constant' $f(x)$. Multiplying the first 
part of the BT (\ref{Liou_BT}) by $e^{\phi_1-\phi_2}$,   
we obtain 
$-(e^{\phi_1-\phi_2})_y = \alpha \, e^{2 \phi_1} = \alpha \, \phi_{1xy}$. 
Integration of the last equation yields
\be
   \phi_2 = \phi_1 - \ln(k - \alpha \, \phi_{1x})  \label{Liou_phi2}
\ee
with integration `constant' $k(x)$. The latter is determined by $\alpha$ 
via 
\be
    (k/\alpha)_x + (1 - k^2) /\alpha^2 = f    \label{Liou_k_alpha}
\ee
which follows with the help of the second BT part in (\ref{Liou_BT}).
Let us now consider two BTs with
\be
  \phi_2 = \phi_1 - \ln(k_{21} - \alpha_{21} \, \phi_{1x}) \, , \qquad
  \phi_3 = \phi_2 - \ln(k_{32} - \alpha_{32} \, \phi_{2x}) \, .
\ee
Eliminating $\phi_2$ from the second equation with the help of the first, 
using (\ref{Liou_phi_xx}) we find
\be
  \phi_3 = \phi_1 - \ln(k_{31} - \alpha_{31} \, \phi_{1x})
\ee
where
\be
 \alpha_{31} = \alpha_{32} k_{21} + k_{32} \, (\alpha_{21} + \alpha_{21x})
    \, , \qquad
 k_{31} = k_{32} k_{21} + \alpha_{32} \, (k_{21x} - \alpha_{21} f_1) 
\ee
solve (\ref{Liou_k_alpha}) with $f_1 =  \phi_{1xx} - \phi_{1x}{}^2$. 
Hence, composition of Liouville BTs preserves their form.

\vskip.1cm
\noindent
{\em Remark.} 
The infinitesimal version of the first of equations (\ref{Liou_BT}) is
$\vari \phi_y = \vari \alpha \, e^{2 \phi + \vari \phi} = \vari \alpha \, e^{2 \phi}$  
where $\vari$ denotes a variation. Using the Liouville equation with $\kappa = 1$, 
this can be integrated with respect to $y$, so that $\vari \phi =  \phi_x \, \vari \alpha - \vari k$ 
where $\vari k(x)$ is an `integration constant'. This is also obtained as the variation of 
(\ref{Liou_phi2}) about $\alpha=0$ and $k=1$.
Together with the variation of the Liouville equation, 
$ \vari \phi_{xy} = 2 \, e^{2 \phi} \, \vari \phi$, we obtain
$\vari k = - {1 \over 2} \vari \alpha_x$ and thus 
$\vari \phi =\vari \alpha \,  \phi_x \,  + {1 \over 2} \vari \alpha_x$.
As a consequence,
$[\vari_1 , \vari_2] \, \phi = \vari \alpha_3 \, \phi_x + {1 \over 2} \, \vari \alpha_{3x}$ 
with $\vari \alpha_3 = \vari \alpha_1 \, \vari \alpha_{2x} - \vari \alpha_2 \, \vari\alpha_{1x}$.

\bigskip
\noindent
{\em 2. Discrete Liouville equation.} Let $M = M^0 \otimes \Lambda_2$ where 
$M^0$ is the set of maps $\mathbb{Z}^2 \rightarrow \mathbb{R}^2$. 
In terms of the shift operators $(S_x z)(x,y) = z(x+1,y)$ and $(S_y z)(x,y) = z(x,y+1)$, 
we define\footnote{Here and in the following we also use the shift operators acting 
on functions via $(S_x \phi)(x,y) = \phi(x+1,y)$.}
\be
  \Dl z = (S_x z - z) \, \xi + (\sigma_+ S_y z - z) \, \tau \, , \quad
  \D z = \kappa \, e^{S_x \phi + \phi} \, \sigma_- S_x z \, \xi
          + \left(e^{ (S_y \phi - \phi) \, \sigma_3}\, S_y z - z \right) \, \tau  
\ee
with a constant $\kappa$. Then $\Dl^2 =0$ and $\D^2 = 0$ identically, and
$\Dl \, \D + \D \, \Dl = 0$ turns out to be equivalent to 
\be
  e^{-\phi(x+1,y)}e^{-\phi(x,y+1)} - e^{-\phi(x+1,y+1)}e^{-\phi(x,y)} = \kappa \, .
\ee
Introducing coordinates $u = (x+y)/\Delta$, $v = (x-y)/\Delta$ with $\Delta^2 = \kappa$ 
and $\phi(x,y) = - \varphi(u-\Delta,v)$, this reads
\be
  e^{\varphi(u,v-\Delta)}e^{\varphi(u,v+\Delta)} 
 - e^{\varphi(u-\Delta,v)}e^{\varphi(u+\Delta,v)} = \Delta^2
\ee
which is Hirota's discretization of the Liouville equation \cite{discr_Liou,Zabr97}.
Let us now explore the corresponding DBTs with $Q^{(0)} = I$. 
Then we have to solve the equation
\be
  [\Dl , Q^{(1)}] = \D_2 - \D_1 
\ee
which restricts $Q^{(1)}$ to the form $Q^{(1)} = r \, \sigma_-$ with 
a function $r$. Furthermore, we obtain the following set of equations,
\be
   r(x,y) &=& e^{\phi_1(x,y) - \phi_1(x,y+1)}
                    - e^{\phi_2(x,y) - \phi_2(x,y+1)}  \label{dLiou_r1} \\
   r(x,y+1) &=& e^{\phi_2(x,y+1) - \phi_2(x,y)}
                    -e^{\phi_1(x,y+1) - \phi_1(x,y)}  \label{dLiou_r2} \\
   r(x+1,y) - r(x,y) &=& \kappa_2 \, e^{\phi_2(x+1,y) + \phi_2(x,y)}
                - \kappa_1 \, e^{\phi_1(x+1,y) + \phi_1(x,y)}  \, . \label{dLiou_r3}
\ee
Using $\pa_{+y} \phi = \phi(x,y+1) - \phi(x,y)$, the first equation can be written as
\be
 r = e^{- \pa_{+y}\phi_1} - e^{-\pa_{+y}\phi_2}
   = (e^{\pa_{+y} \phi_2} - e^{\pa_{+y}\phi_1}) \, e^{- \pa_{+y}(\phi_1+\phi_2)}
\ee
with the help of which we can convert (\ref{dLiou_r2}) into
\be
   S_y r = r \, e^{\pa_{+y}(\phi_1+\phi_2)}  \; .
\ee
This equation can be `integrated' and yields
\be
   r = \alpha(x) \, e^{\phi_1 + \phi_2}  \label{dLiou_r}
\ee
with an arbitrary function $\alpha(x)$. Together with (\ref{dLiou_r1}), 
it leads to the first part of the BT,
\be
  e^{- S_y \phi_1} \, e^{-\phi_2}- e^{-\phi_1} \, e^{- S_y \phi_2} = \alpha \, .
\ee
The other BT part follows from (\ref{dLiou_r3}):
\be
      (S_x \alpha) \, e^{S_x \phi_1 + S_x \phi_2} - \alpha \, e^{\phi_1 + \phi_2} 
  =  \kappa_2 \, e^{S_x \phi_2 + \phi_2} - \kappa_1 \, e^{S_x \phi_1 + \phi_1} \, .
\ee
 For $N=1$, no additional equations arise from the remaining DBT conditions. 
 For $N>1$, we could at most get restrictions on the above BT. In particular, 
no new BTs can show up, as in the case of the continuous Liouville equation.
Taking (\ref{dLiou_r}) into account, the $N=1$ expression for $Q$ is the same 
as for the continuum model. Hence we obtain the same permutability condition.

\section{Darboux-B\"acklund transformations of dressed bicomplexes}
\setcounter{equation}{0}
All the examples presented in the next section possess a somewhat more 
specialized form of the bicomplex equations than what is allowed by the 
general formalism of section 1. For this class one can make some general 
observations which help to reduce the amount of calculations needed to 
elaborate the DBTs in concrete examples. The corresponding formalism 
is developed in this section. Clearly, this is of a more technical nature. 
In principle, given a bicomplex formulation of some equation, the formulas 
of section 1 are sufficient to work out the corresponding DBTs. In a 
given example, however, it may turn out to be very difficult to do it in a 
straight way.
\vskip.1cm

It is often convenient \cite{DMH-bico,DMH-bidiff} to start with a 
trivial\footnote{`Trivial' in the sense that the corresponding bicomplex 
conditions are identically satisfied.}  
bicomplex and to use what we call `dressings' to construct non-trivial 
bicomplexes. Normally, such a `deformation' of a trivial bicomplex results 
in too many independent equations, but there are two particular ways of 
introducing dressings (see (A) and (B) below) which keep some of the 
bicomplex equations identically satisfied.
\vskip.1cm

Let $(M, \dl, \d)$ be a trivial
bicomplex and $L$ the space of linear operator valued 
forms\footnote{The elements of $M^s$ are called $s$-forms.} 
acting on $M$, i.e., for $z \in M$ and $T \in L$ we have $Tz \in M$. 
On operators we define
\be
   \td T = [\d,T] \, ,  \qquad \qquad
   \tdl T = [\dl,T]
\ee
where $[ \, , \, ]$ is the graded commutator.\footnote{$\dl$ and $\d$ are odd. 
 For an even operator $T$, $[\d,T] = \d T - T \d$. For an odd $T$, 
$[\d,T] = \d T + T \d$.}  
Then $(L,\tdl,\td)$ is again a bicomplex and, moreover, a bi-differential 
calculus\footnote{Besides $\tdl^2 = \td^2 = \td \tdl + \tdl \td = 0$, 
$\tdl$ and $\td$ obey the Leibniz rule, i.e., the graded product rule of 
differentiation.}    
\cite{DMH-bidiff}. A dressing of the bicomplex $(M,\dl,\d)$ is a new 
bicomplex $(M,\Dl,\D)$ where
\be
   \D z = \d z + A \, z \, , \qquad  \Dl z = \dl z + B \, z  \label{dressed}
\ee
with `connection' 1-forms $A,B \in L$. The conditions for $(M,\Dl,\D)$ to be 
a bicomplex impose the following conditions on $A$ and $B$,
\be
   F = \td A + A^2 = 0 \, , \qquad {\cal F} = \tdl B + B^2=0 \, , \qquad 
  \tdl A + \td B + A \, B + B A = 0 \, .         \label{bic_AB}
\ee
Introducing a real parameter $\lambda$ and
\be
   \d_\lambda = \dl - \lambda \, \d \, , \qquad 
   A(\lambda) = B - \lambda \, A \, ,
\ee
the three conditions (\ref{bic_AB}) can be compactly written as a 
$\lambda$-dependent zero curvature condition, 
\be
   F(\lambda) = \td_\lambda A(\lambda) + A(\lambda)^2 = 0
\ee
for all $\lambda$.
\vskip.1cm

Now we consider two dressings with $A_i, B_i$, $i=1,2$, and look for 
a DBT of the two resulting bicomplexes ${\cal B}_i$.\footnote{Note that 
in this framework the linear spaces $M_i$ of the two bicomplexes 
are taken to be the same. If $Q$ is invertible, this is not a restriction as 
pointed out in section 1.} 
The DBT condition (\ref{DBT}) then becomes
\be
  \td_\lambda Q + A_2(\lambda) \, Q 
    - Q \, A_1(\lambda) = 0 \, .
\ee
In terms of $A$ and $B$, this reads
\be
  \tdl Q + B_2 \, Q - Q \, B_1  = \lambda \,  \tD_{21} Q
                                    \label{Q_AB}
\ee
where
\be
   \tD_{21} Q = \td Q + A_2 \, Q - Q \, A_1 \; .
\ee
In the following we confine our considerations to the case where 
$B_1 = 0 = B_2$.\footnote{Actually, in general, this can be achieved by 
separate gauge transformations and coordinate transformations 
of the two bicomplexes, since ${\cal F}_i = 0$. See section 5 for an 
example.} 
Then the conditions (\ref{bic_AB}) for the two bicomplexes reduce to
\be
   F_i = \td A_i + A_i^2 = 0 \, , \qquad \qquad 
   \tdl A_i = 0            \label{bic_eq_i}
\ee
and (\ref{Q_AB}) becomes
\be
     \tdl Q = \lambda \, \tD_{21} Q \, . \label{lin}
\ee
This equation has to be solved in order to determine the DBTs of a dressed 
bicomplex. Using the ansatz (\ref{Q_Nth}), the DBT condition splits into the 
following set of equations,
\be
    \tdl Q^{(0)} = 0 \, , \quad  
    \tdl Q^{(k)} = \tD_{21} Q^{(k-1)} \quad (k=1, \ldots, N) \, , \quad
    \tD_{21} Q^{(N)} = 0  \, .    \label{Nth_BT}
\ee
\vskip.1cm

Now we assume that $Q$ is invertible with\footnote{This choice for  $Q^{(0)}$ 
trivially solves the first of equations (\ref{Nth_BT}). In the case of auto-DBTs, 
arguments for this choice (under certain assumptions) have been given in section 1.}
 $Q^{(0)} = I$ and consider in more detail the case of a primary DBT with 
\be
      Q = I + \lambda \, R   \label{Q_linear}
\ee
where $R = Q^{(1)}$ does not depend on $\lambda$. 
(\ref{Nth_BT}) then reduces to 
\be
  \tdl R = \tD_{21} I = A_2 - A_1 \, , \qquad 
  \tD_{21} R = 0  \, .  \label{primaryDBT}
\ee
In fact, in all the examples which we explored so far, it turned out to be sufficient 
to consider such a primary invertible DBT in order to recover 
well-known BTs. We can somewhat simplify the last set of equations as follows, 
using two obvious ways to reduce the set of bicomplex equations 
\cite{DMH-bico}.\footnote{The following is specific to bicomplexes 
and has no analogue in case of a zero curvature condition non-linear in 
$\lambda$. Though such a condition can also be solved as in (A) 
by writing the $\lambda$-dependent gauge potential as a ``pure gauge",  
the corresponding $g$ then depends on $\lambda$.}

\bigskip\noindent
{\bf (A)} Let $A_i = g_i^{-1} \td g_i$ where $g_i : M^0 \rightarrow M^0$ 
are invertible operators (e.g., matrices) not depending on $\lambda$. 
This solves the first of the bicomplex equations (\ref{bic_eq_i}). 
The second of equations (\ref{primaryDBT}) is then 
equivalent to $\td (g_2 R  g_1^{-1}) = 0$ which can be converted to 
$\td a = 0$ by setting $R  = g_2^{-1} \, a \, g_1$ with $a \in L^0$.\footnote{For 
arbitrary $N$, this holds with $R$ replaced by $Q^{(N)}$.}
The first of equations (\ref{primaryDBT}) now becomes
\be
   \tdl (g_2^{-1} a \, g_1) 
 = g_2^{-1} \td g_2 - g_1^{-1} \td g_1 
 = g_2^{-1} \td(g_2 g_1^{-1}) \, g_1 \, .   \label{gba}
\ee

\medskip\noindent
{\bf (B)} Let $A_i = \tdl w_i$ with $w_i \in L^0$ not depending on $\lambda$. 
This solves the second of the bicomplex equations (\ref{bic_eq_i}). From the 
first equation of (\ref{primaryDBT}) we obtain 
\be
       R = w_2 - w_1 + T    \label{R_schemeB}
\ee 
with $\tdl T = 0$.
Then the second equation in (\ref{primaryDBT}) becomes
\be
  \Phi_2 - \Phi_1 + \td T + \tdl \left( w_2 T - T w_1 
  + {1 \over 2} (w_2-w_1)^2 + {1 \over 2} [w_1 , w_2] \right) = 0 
          \label{BT_B1}
\ee
where 
\be
    \Phi_i = \td w_i + {1 \over 2} [\tdl w_i , w_i]
\ee
are $\tdl$-potentials of the curvatures $F_i$, i.e., 
\be
     \tdl \Phi_i = -F_i \; . 
\ee
Let us assume that the first $\tdl$-cohomology is trivial. The last equation 
together with $F_i = 0$ then implies
\be
    \Phi_i = \tdl \Psi_i  \; .
\ee
 Furthermore, $\tdl \td T = 0$ leads to $\td T = \tdl b$ with $b \in L^0$. 
Then (\ref{BT_B1}) can be integrated and leads to\footnote{An `integration 
constant' $c$ with $\tdl c = 0$ can be absorbed by a redefinition of $b$.} 
\be
    \Psi_2 - \Psi_1 + w_2 T - T w_1 
 + {1 \over 2} (w_2-w_1)^2 + {1 \over 2} [w_1 , w_2] + b = 0 \; .  
\ee 
In concrete examples, it is often simpler to work out directly the second of 
equations (\ref{primaryDBT}), however.
\vskip.1cm

 For a DBT with $Q$ of the form (\ref{Q_Nth}) where $Q^{(0)} = I$, 
the permutability condition (\ref{permutability}) results in the following system 
of equations:
\be
     Q^{(1)}_{31} + Q^{(1)}_{10} - Q^{(1)}_{32} - Q^{(1)}_{20}   
 &=& 0 \\
     Q^{(k)}_{31} + Q^{(k)}_{10} - Q^{(k)}_{32} - Q^{(k)}_{20}   
 &=&  \sum_{m=1}^{k-1} \left( Q^{(m)}_{32} Q^{(k-m)}_{20} 
      - Q^{(m)}_{31} Q^{(k-m)}_{10} \right)   
     \quad k=2, \ldots, N  \qquad \quad \\
   \sum_{m+n=k \atop 1 \leq m,n \leq N} \left( Q^{(m)}_{31} Q^{(n)}_{10} 
     - Q^{(m)}_{32} Q^{(n)}_{20} \right)  
 &=& 0  \qquad  k = N+1, \ldots, 2N \, .  
\ee
 For a primary DBT with $Q = I + \lambda \, R$, this reduces to 
\be
   R_{31} + R_{10} &=& R_{32} + R_{20}   \label{perm1}  \\
   R_{31} \, R_{10} &=& R_{32} \, R_{20}  \, .  \label{perm2}
\ee
In case (A), we have 
$R_{ij} = g_i^{-1} a_{ij} \, g_j$, so that (\ref{perm2}) becomes
\be
    a_{31} \, a_{10} = a_{32} \, a_{20}   \, .  \label{perm2A}
\ee
In case (B), we have $R_{ij} = \phi_i - \phi_j + T_{ij}$. 
Then (\ref{perm1}) reduces to 
\be
    T_{31} + T_{10} = T_{32} + T_{20} \; .
\ee

\section{Bicomplexes and auto-Darboux-B\"acklund transformations for various 
         integrable models}
\setcounter{equation}{0}
In the following, we elaborate auto-DBTs for various integrable models. 
All the examples of non-trivial bicomplexes in this section are of the form 
$(M,\dl,\D)$ where $(M,\dl,\d)$ is a trivial bicomplex and $\D$ has the 
decomposed `dressed' form $\D = \d + A$ as considered in the previous 
section. 
Assuming $Q$ to be invertible, in section 1 we have motivated a  
restriction of the auto-DBT condition to the form (\ref{autoDBT}) 
with $Q^{(0)} = I$. This is the basis for the following calculations. 
As in section 2, $\Lambda_2$ denotes the exterior algebra of a 
two-dimensional real vector space.

\subsection{KdV and related equations}
\bigskip
\noindent
{\em 1. KdV equation, primary DBT.} 
Let $M = C^\infty(\mathbb{R}^2, \mathbb{R}) \otimes \Lambda_2$ with 
\be
   \dl z = -3 z_{xx} \, \tau + z_x \, \xi \, , \qquad
   \d z  = (z_t + 4 z_{xxx}) \, \tau - z_{xx} \, \xi  
                            \label{delta_d_KdV}
\ee
for $z \in M^0$. Dressed with the gauge potential 1-form
\be
   A = \tdl w = - 3 \, (w_{xx} + 2w_x \, \pa_x) \, \tau + w_x \, \xi
                              \label{A_KdV}
\ee
we get a bicomplex for the KdV equation
\be
    u_t + u_{xxx} - 6 u u_x  = 0
\ee
where $u = w_x$ (see also \cite{DMH-KdV}). Here we choose $w \in L^0$ as 
a function (which acts by multiplication). Looking for a primary 
DBT, we have (\ref{R_schemeB}) with $T_x = 0$. Furthermore, 
\be
      \tD_{21} R 
  &=& \{ R_t + 4 R_{xxx} + 12 R_{xx} \, \pa_x + 12 R_x \, \pa_x^2
      -3 \, [(w_{2xx} + 2 w_{2x} \, \pa_x) \, R     \nonumber \\
  & & - R \, (w_{1xx} + 2 w_{1x} \, \pa_x)] \} \, \tau 
      - (R_{xx} + 2 R_x \, \pa_x - w_{2x} \, R + R \, w_{1x}) \, \xi 
               \label{DR_KdV}
\ee
has to vanish. Using (\ref{R_schemeB}), the $\xi$-part leads to 
\be
   (w_2 - w_1)_{xx} + 2 (w_2 - w_1)_x \, \pa_x - w_{2x} \, (w_2 - w_1 + T) 
   + (w_2 - w_1 + T) \, w_{1x} = 0 \; .    \label{KdV_w2-w1}
\ee
In particular, this implies $T = 2 \pa_x + \beta$ with a function $\beta(t)$ which, however, 
can be absorbed via a redefinition of $w_1$ (which leaves the KdV equation invariant). 
Hence
\be
   R = w_2 - w_1 + 2 \pa_x      \label{KdV-R}
\ee
and (\ref{KdV_w2-w1}) leads to the BT part
\be
  (w_1 + w_2)_x = 2 \alpha + {1 \over 2} \, (w_2 - w_1)^2  
                                    \label{kdv1}
\ee
with an integration `constant' $\alpha$ which is an arbitrary function of $t$. 
The vanishing of the $\tau$-part of (\ref{DR_KdV}) together with (\ref{kdv1}) 
yields\footnote{This is also obtained by integration of the difference of the two 
KdV equations for $u_1 = w_{1x}$ and $u_2 = w_{2x}$ with vanishing 
integration `constant'.}
\be
   (w_2 - w_1)_t + (w_2 - w_1)_{xxx} 
    - 3 (w_2 - w_1)_x \, (w_1 + w_2)_x = 0  
\ee
which is the second BT part for the KdV equation (see \cite{Draz+John89}, 
p.113, for example, and also \cite{BT,Roge+Shad82,BT-KdV,Lamb74}). 
Introducing 
\be
     r = w_2 - w_1
\ee
the BT can also be written as
\be
  w_{1x} = \alpha - {1 \over 2} \, r_x + {1 \over 4} \, r^2  \, , \quad
     r_t + [ r_{xx} - {1 \over 2} r^3 - 6 \alpha r ]_x = 0  \; .  
\ee
The last equation has the form of a conservation law and can be integrated once 
if we set 
\be
    r = - 2 \, (\ln \chi)_x  \; .
\ee
Then, in terms of $\chi$, the BT reads
\be
 w_{1x} = \alpha + {\chi_{xx} \over \chi} \, , \qquad
 {\chi_t \over \chi_x} = 6 \alpha - {\chi_{xxx} \over \chi_x} 
 + 3 {\chi_{xx} \over \chi} + \gamma {\chi \over \chi_x} 
           \label{KdV_BT_chi}
\ee
with a `constant of integration' $\gamma(t)$, assuming $\chi_x \neq 0$. 

The following will be needed below for our discussion of the secondary DBT. 
We consider two BTs,
\be
   (w_1 + w_2)_x = 2 \alpha_1 + {1 \over 2} (w_1 - w_2)^2  \, ,  
                   \quad
   (w_2 + w_3)_x = 2 \alpha_2 + {1 \over 2} (w_2 - w_3)^2 \, .
                               \label{KdV_2BTs}
\ee
The difference of both equations is
\be
  (w_1 - w_3)_x = 2 (\alpha_1 - \alpha_2) 
                  + {1 \over 2} (w_1 - w_3) (w_1 + w_3 - 2w_2) \, .
\ee
Solving for $w_2$ yields
\be
  w_2 = {1 \over 2} (w_1 + w_3) - {s_x \over s} 
        + {2 (\alpha_1 - \alpha_2) \over s} \, , \qquad
   s = w_1 - w_3  \; .
\ee
Inserting this expression into the sum of the two equations 
(\ref{KdV_2BTs}), we obtain
\be
  (w_1 + w_3)_x = \alpha_1 + \alpha_2 + {s_{xx} \over s} 
   - {1 \over 2} \left( {s_x \over s} \right)^2 
   + {1 \over 8} s^2 + {2 (\alpha_1 - \alpha_2)^2 \over s^2} \, . 
                         \label{w1+w3}
\ee
The complementary parts of the above two BTs are
\be
  (w_1 - w_2)_t + (w_1-w_2)_{xxx} - 3 \, (w_{1x}^2 - w_{2x}^2) &=& 0 \, , \\
  (w_2 - w_3)_t + (w_2-w_3)_{xxx} - 3 \, (w_{2x}^2 - w_{3x}^2) &=& 0 \, .
\ee
Adding these two equations leads to
\be
   s_t + s_{xxx} - 3 s_x \, (w_1 + w_3)_x = 0  \, .
\ee
Using (\ref{w1+w3}) to eliminate $(w_1+w_3)_x$, we obtain
\be
   s_t + [ s_{xx} - 3 (\alpha_1 + \alpha_2) \, s - {1 \over 8} s^3 
   + 6 (\alpha_1 + \alpha_2)^2 {1 \over s} - {3 s_x^2 \over 2 s} ]_x = 0 
\ee
which, setting $s = - 2 (\ln \chi)_x$ and integrating once with 
integration `constant' $2 \epsilon(t)$, becomes
\be
  {\chi_t \over \chi_x} = 3 (\alpha_1 + \alpha_2) - {\cal S}_x \chi 
 - {3 (\alpha_1 - \alpha_2)^2 \over 2}
  \left( {\chi \over \chi_x} \right)^2 + \epsilon {\chi \over \chi_x} \, .
\ee
Here, ${\cal S}_x$ denotes the Schwarzian derivative
\be  
  {\cal S}_x \chi = {\chi_{xxx} \over \chi_x}
   - {3 \over 2} \, \left( {\chi_{xx} \over \chi_x} \right)^2 \; .
                 \label{Schw_deriv}
\ee
Correspondingly, (\ref{w1+w3}) can be rewritten 
in the form
\be
  w_{3x} = {\alpha_1 + \alpha_2 \over 2} + {\chi_{xxx} \over 2 \chi_x}
          - \left( {\chi_{xx}{} \over 2 \chi_x} \right)^2 
          + {(\alpha_1 - \alpha_2)^2 \over 4} \left( {\chi \over \chi_x} \right)^2 
          \, .
\ee
 From these expressions one recovers for $\alpha_1 = \alpha_2$ and $\epsilon =0$ 
a BT found by Galas \cite{Galas}.
The latter is therefore just the composition of two `elementary' BTs.

\bigskip
\noindent
{\em 2. KdV equation, permutability.} With (\ref{KdV-R}), the permutability condition 
(\ref{perm1}) is identically satisfied and (\ref{perm2}) leads to
\be
    w_3 = - w_0 + w_1 + w_2 
      - {2 \, ( w_1 - w_2 )_x \over w_1 - w_2 }  \; .  
                           \label{KdV_perm1}
\ee
With the help of (\ref{kdv1}), for the pairs $w_0,w_1$ with $\alpha_1$ and 
$w_0,w_2$ with $\alpha_2$, the last equation can be written as
\be
   w_3 = w_0 - 4 \, {\alpha_1 - \alpha_2 \over w_1 - w_2}  \, .
\ee

\bigskip
\noindent
{\em 3. KdV equation, secondary DBT.} 
We consider again the above bicomplex associated with the KdV equation. 
But now we turn to the secondary DBT. Again, we have 
$Q^{(1)} = w_2 - w_1 + T$ with $T_x = 0$. For $N=2$, (\ref{Nth_BT}) 
then requires
\be
   \tdl Q^{(2)} &=& \td ( w_2 - w_1 + T ) + (\tdl w_2) \, w_2 + w_1 \, \tdl w_1 
        + \tdl ( - w_1 w_2 + w_2 T - T w_1 )  
                            \label{KdV_BT2a}  \\
   \td Q^{(2)} &=& Q^{(2)} \, \tdl w_1 - (\tdl w_2) \, Q^{(2)} \; .
                            \label{KdV_BT2b}
\ee
The $\xi$-part of (\ref{KdV_BT2a}) can be integrated and leads to 
\be
   Q^{(2)} = (w_1 - w_2)_x + {1 \over 2} ( w_1 - w_2 )^2 
             + 2 (w_1 - w_2) \, \pa_x + w_2 T - T w_1 + \rho
\ee
where $\rho_x = 0$. Inserted in the $\xi$-part of (\ref{KdV_BT2b}), 
which is
\be
   Q^{(2)}_{xx} + Q^{(2)} \, w_{1x} - w_{2x} \, Q^{(2)} 
 + 2 \, Q^{(2)}_x \, \pa_x = 0 \, ,
\ee
this enforces $T = 4 \pa_x + \beta(t)$. The function $\beta(t)$ can 
be absorbed by a redefinition of $w_1$. Furthermore, we obtain 
\be
   Q^{(2)} = - r_x + {1 \over 2} r^2 - 4 \, w_{1x} + 2 r \, \pa_x 
             + 4 \, \pa_x^2 + 4 \alpha
\ee
with $r = w_2 - w_1$ and an arbitrary function $\alpha(t)$,  
and finally
\be
   (w_1 + w_2)_x = 2 \alpha + {r_{xx} \over r} 
 - {1 \over 2} \left( {r_x \over r} \right)^2 
 + {1 \over 8} r^2 - {8 \gamma \over r^2}   \label{KdV_BT2_1}
\ee
with an integration `constant' $\gamma(t)$.
The $\tau$-part of (\ref{KdV_BT2a}) is now evaluated to
\be
   r_t + r_{xxx} - 3 r_x \, (r_x + 2 w_{1x}) = 0 \, .                            
\ee
Elimination of $w_{1x}$ with the help of (\ref{KdV_BT2_1}) 
leads to
\be
    r_t + [ r_{xx} - {3 \over 2} {r_x^2 \over r} - 6 \alpha \, r 
  - {1 \over 8} r^3 - 24 {\gamma \over r} ]_x = 0 \, .               
\ee
Setting $r = - 2 (\ln \chi)_x$, this equation can be integrated 
and rewritten as 
\be
  {\chi_t \over \chi_x} = 6 \alpha - {\cal S}_x \chi 
 + 6 \gamma \left( {\chi \over \chi_x} \right)^2 
 + \epsilon {\chi \over \chi_x} 
\ee
with an arbitrary function $\epsilon(t)$ and the Schwarzian derivative 
defined in (\ref{Schw_deriv}). 
 In terms of $\chi$, (\ref{KdV_BT2_1}) takes the form
\be
   w_{1x} = \alpha + {\chi_{xxx} \over 2 \chi_x} 
     - \left( {\chi_{xx} \over 2 \chi_x} \right)^2 
     - \gamma \left( {\chi \over \chi_x} \right)^2  \, .
\ee
 Comparison with our previous results shows 
that this secondary DBT for the KdV bicomplex is just 
the composition of two primary DBTs. If $\gamma < 0$, this 
is evident. If $\gamma > 0$ we set 
$\alpha_1 = \alpha + i \sqrt{\gamma}$ and 
$\alpha_2 = \alpha - i \sqrt{\gamma}$. Although in this case 
the elementary BTs (with $\alpha_1$ and $\alpha_2$, respectively) 
do not produce {\em real} solutions from real solutions, in general, 
their composition does.

\bigskip
\noindent
{\em 4. Modified KdV equation.}
Supplied with the product $(a,b)(a',b') = (a a',b b')$, the vector space $\mathbb{R}^2$ 
becomes a commutative ring with unit $(1,1)$ which we denote as ${}^2\mathbb{R}$ 
\cite{Port69}. It is a realization of the abstract commutative ring generated by a unit $1$ 
and another element ${\bf e}$ satisfying ${\bf e}^2=1$. Here we have 
${\bf e} = (1,-1)$. 
The relevance of ${}^2\mathbb{R}$ for  the mKdV equation stems from the
following observation (see also \cite{Ivan+Popo92}). Let $u$ be a field with values 
in ${}^2\mathbb{R}$ which satisfies the KdV equation
\be
     u_t + u_{xxx} - 6 u u_x = 0  
\ee
and let
\be
   u = v^2 + v_x \, {\bf e}                \label{Miura}
\ee
with a real valued field $v$. Now
\be
   u_t + u_{xxx} - 6 u u_x = (2 v + {\bf e} \, \pa_x) (v_t + v_{xxx} - 6 v^2 v_x)
\ee
shows that the field $v$ satisfies the mKdV equation
\be
    v_t + v_{xxx} - 6 v^2 v_x = 0  \, .
\ee
In fact, (\ref{Miura}) is the famous Miura transformation and its `conjugate'
since with $u = (u^+,u^-) = (v^2,v^2) + (v_x,-v_x)$ we obtain
\be
  u^+ = v_x + v^2 \, ,  \qquad
  u^- = - v_x + v^2  \, .
\ee
Consequently, the two KdV equations for $u^\pm$ are equivalent to the
above mKdV equation. Hence, in order to find an auto-BT for the mKdV 
equation, we simply have to extend our KdV treatment to fields with values 
in ${}^2\mathbb{R}$, though we have to take care of the fact that ${}^2\mathbb{R}$ 
is not a division ring (since $(1,0)(0,1) = (1+{\bf e})(1-{\bf e}) = 0$). 
Introducing $w = \hat{v}+ v \, {\bf e}$ with $\hat{v}_x = v^2$, we have $w_x = u$ 
and we can directly generalize the KdV auto-BT:  
\be
  (w_1 + w_2)_x  - 2 \alpha   - {1 \over 2} \, (w_1 - w_2)^2 &=& 0
                                    \label{mkdv1}  \\
   (w_1 - w_2)_t + (w_1 - w_2)_{xxx}
    - 3 (w_1 - w_2)_x \, (w_1 + w_2)_x &=& 0      \label{mkdv2}
\ee
(where $\beta(t)$ has been absorbed in $w_1$). Note
that the integration `constant' $\alpha(t)$  is now an element 
of ${}^2\mathbb{R}$.
We decompose (\ref{mkdv1}) with $\alpha = - k^2 + b \, {\bf e}$ 
to obtain the two equations
\be
  (\hat{v}_1 + \hat{v}_2)_x &=& -2 k^2 + {1 \over 2} [(\hat{v}_1 - \hat{v}_2)^2
   + (v_1-v_2)^2]         \label{m1a}  \\
  (v_1 + v_2)_x &=& 2 b + (\hat{v}_1 - \hat{v}_2) (v_1 - v_2) \, .  \label{m1b}
\ee
Using $\hat{v}_x = v^2$ in (\ref{m1a}), we obtain
\be
  (v_1 + v_2)^2 = - 4 k^2 + (\hat{v}_1 - \hat{v}_2)^2  \, .
\ee
Applying $\pa_x$ to this equation and comparing the result with (\ref{m1b}), 
we find $b=0$. Solving the last equation for $\hat{v}_1 - \hat{v}_2$ 
and inserting this expression in (\ref{m1b}), leads to the first part of the 
auto-BT,\footnote{With $v = q_x $, this reads 
$(q_1 + q_2)_x = 2 k \, \sinh(q_1 - q_2) $ where an integration constant 
has been absorbed by a redefinition of the $q_i$.}
\be
    (v_1 + v_2)_x = \pm (v_1 - v_2) \, \sqrt{4 k^2 + (v_1+v_2)^2}  \, .
\ee
 Decomposition of (\ref{mkdv2}) leads to 
\be
  (\hat{v}_1 - \hat{v}_2)_t + (\hat{v}_1 - \hat{v}_2)_{xxx} - 3 \, (\hat{v}_1 - \hat{v}_2)_x \, 
  (\hat{v}_1 + \hat{v}_2)_x - 3 \, (v_1 - v_2)_x \, (v_1 + v_2)_x &=& 0 \\
  (v_1 - v_2)_t + (v_1 - v_2)_{xxx} - 3 (\hat{v}_1 - \hat{v}_2)_x \, (v_1 + v_2)_x 
  - 3 \, (v_1 - v_2)_x \, (\hat{v}_1 + \hat{v}_2)_x &=& 0 \, .
\ee
Using $\hat{v}_x = v^2$ in the second equation produces the second part of the 
mKdV auto-BT, 
\be
   (v_1 - v_2)_t + [(v_1 - v_2)_{xx} - 2 \, v_1^3 + 2 \, v_2^3]_x = 0 
\ee
which is the difference of the two mKdV equations for $v_1$ and $v_2$.

\bigskip
\noindent
{\em 5. mKdV equation, permutability.}
In the framework of the previous subsection, the KdV permutability 
condition takes the form
\be
   (w_2 - w_1) (w_1 + w_2 - w_0 - w_3) = 2 \, (w_2 - w_1)_x
\ee
(cf (\ref{KdV_perm1})) and, by use of (\ref{mkdv1}),
\be
   (w_3 - w_0) (w_2 - w_1) =  4 \, (k_2^2 - k_1^2) 
\ee
from which we obtain, by decomposition,
\be
   (v_3 - v_0) (\hat{v}_2 - \hat{v}_1) 
 + (v_2 - v_1) (\hat{v}_3 - \hat{v}_0) &=& 0 \\
   (\hat{v}_3 - \hat{v}_0)(\hat{v}_2 - \hat{v}_1)
  + (v_3 - v_0) (v_2 - v_1) &=& 4 \, (k_2^2 - k_1^2)
\ee
and thus the following superposition formula for mKdV 
solutions, 
\be
  v_3 = v_0 + { 4 (k^2_2 - k^2_1) (v_2-v_1) \over
   (v_2 - v_1)^2 - (\hat{v}_2 - \hat{v}_1)^2 }  \, .
\ee

\bigskip
\noindent
{\em 6. ncKdV equation.} 
We choose the bicomplex maps and the dressing as for the KdV 
equation, so that (\ref{delta_d_KdV}) and (\ref{A_KdV}) hold.
But now we take $u$ to be a map from $\mathbb{R}^2$ into 
some noncommutative associative algebra with product $\ast$ 
for which $\pa_t$ and $\pa_x$ are derivations. 
Then we have a bicomplex iff $\td A + A \ast A = 0$ which is 
equivalent to the {\em noncommutative KdV equation} (ncKdV) 
\be
   u_t + u_{xxx} - 3 (u \ast u_x + u_x \ast u) = 0 
\ee
where $u = w_x$ \cite{DMH-ncKdV}. The corresponding potential 
ncKdV equation is
\be
   w_t + w_{xxx} - 3 w_x \ast w_x = 0 \, .
\ee
The conditions for a primary DBT with $Q$ of the form (\ref{Q_linear}) 
are
\be
  \tdl R = A_2 - A_1 \, , \qquad   
  \td R + A_2 \ast R - R \ast A_1 = 0  \, .
\ee
As in the ordinary KdV case, the first equation is solved by 
$R = w_2 - w_1 + T$ with $T_x=0$. The second equation
implies $T = 2 \pa_x + \beta(t)$. Again, the function $\beta(t)$ 
expresses the freedom in the choice of the potential for $u$ and 
can be set to zero. Furthermore, we obtain
\be
  (w_1+w_2)_{xx} - w_1 \ast w_{1x} - w_{2x} \ast w_2  + (w_2 \ast w_1)_x 
       &=& 0          \label{astBT1}   \\
  (w_2-w_1)_t + (w_2-w_1)_{xxx} - 3 (w_{2x} \ast w_{2x} - w_{1x} \ast w_{1x}) 
       &=& 0 \, .     \label{astBT2}
\ee
In the commutative case, (\ref{astBT1}) can be integrated (which introduces 
a parameter in the BT). This is not so in the noncommutative case. But 
we still have a BT which is no longer symmetric in $w_1$ and $w_2$, 
however. 

\bigskip
\noindent
{\em 7. ncKdV equation, permutability.} 
The permutability conditions are reduced to 
$R_{31} \ast R_{10} = R_{32} \ast R_{20}$. This yields
\be
  w_3 \ast (w_1-w_2) + (w_1-w_2) \ast w_0 = w_1 \ast w_1 - w_2 \ast w_2 
  - 2 (w_1-w_2)_x  
\ee
where each of the pairs $(w_0,w_1)$, $(w_0,w_2)$, $(w_1,w_3)$, $(w_2,w_3)$ 
has to satisfy the BT equations (\ref{astBT1}) and (\ref{astBT2}).
 Let us now specify the $\ast$-product as the Moyal product 
\be
   f \ast h = {\bf m} \circ e^{\vt P/2} (f \otimes h) 
\ee
for smooth functions $f,h$, where ${\bf m}(f \otimes h) = f h$, 
$P = \pa_t \otimes \pa_x - \pa_x \otimes \pa_t$, and $\vt$ is a deformation 
parameter. As an example, let $w_3 = 0$, so that
\be
   (w_1 - w_2) \ast w_0  = w_1 \ast w_1 - w_2 \ast w_2 - 2 (w_1 - w_2)_x \; .
            \label{ncKdV-perm}
\ee
 For $w_1$ and $w_2$ we choose the 1-soliton solutions
\be 
   w_1 = - 2 \, \mbox{tanh}(x - 4 t) \, , \quad 
   w_2 = - 4 \, \mbox{coth}(2 x - 32 t)  
\ee
(see also \cite{Draz+John89}, p.116). $(w_1,w_3)$ and $(w_2,w_3)$ indeed 
satisfy (\ref{astBT1}) and (\ref{astBT2}). 
Then the $\ast$-products on the right side of (\ref{ncKdV-perm}) reduce to 
ordinary products and we obtain
\be  
       f(x,t) \ast w_0 = g(x,t)
\ee
with
\be
 g(x,t) &=& 4 \, \mbox{tanh}^2(x - 4 t) 
            - 16 \, \mbox{coth}^2(2 x - 32 t) - 2 \, f_x  \\        
 f(x,t) &=& - 2 \, \mbox{tanh}(x - 4 t) 
            + 4 \, \mbox{coth}(2 x - 32 t) \; .
\ee
This implies
\be
    0 = {\pa \over \pa \vt} (f \ast w_0) 
      = f \ast {\pa w_0 \over \pa \vt}  
        + {1 \over 2} ( f_t \ast w_{0x} - f_x \ast w_{0t} ) 
\ee
and
\be
   f \ast {\pa^2 w_0 \over \pa \vt^2}
 = - [ f_t \ast {\pa w_{0x} \over \pa \vt} 
   - f_x \ast {\pa w_{0t} \over \pa \vt} ] 
   - {1 \over 4} ( f_{tt} \ast w_{0xx} - 2 f_{tx} \ast w_{0xt} 
   + f_{xx} \ast w_{0tt} ) \; .
\ee
 For vanishing deformation parameter $\vt$, the solution of the 
permutability conditions is the 2-soliton solution
\be
    W_0 = - 6 /[ 2 \, \coth (2 \, x - 32 \, t) - \tanh (x - 4 \, t) ] 
\ee
with corresponding KdV solution
\be
  u_0 = W_{0x} = -12 \, \frac{3 + \cosh (4 \, x - 64 \, t) 
        + 4 \, \cosh (2 \, x - 8 \, t) }{ \left( \cosh
        (3 \, x - 36 \, t) + 3 \, \cosh (x - 28 \, t)
        \right)^2}     \label{KdV-2s}
\ee
(cf \cite{Draz+John89}, p.116). The noncommutative solution is then 
\be
     w_0 = W_0 + \vt \, W_1 + {1 \over 2} \vt^2 \, W_2 + \ldots
\ee
where
\be
     W_1
  &:=& \left( {\pa w_0 \over \pa \vt} \right)_{\vt = 0} 
   = - {1 \over 2 f} ( f_t \, W_{0x} - f_x \, W_{0t} )   \\
      W_2
  &:=& \left( {\pa^2 w_0 \over \pa \vt^2} \right)_{\vt = 0} 
    = - {1 \over f} ( f_t \, W_{1x} - f_x \, W_{1t} )  
      - {1 \over 4 f} ( f_{tt} \, W_{0xx} - 2 f_{tx} \, W_{0xt} 
      + f_{xx} \, W_{0tt} )   \qquad
\ee
and so forth. In the case under consideration, we obtain $W_1 = 0$ and
\be
  W_{2x} = 331776 \,
  \frac{\left( \cosh (3 \, x - 36 \, t) - 3 \, \cosh (x - 28 \, t) \right) 
  \, \left( \sinh (3 \, x - 36 \, t) + \sinh (x - 28 \, t) \right)^2}
  {\left( \cosh(3 \, x - 36 \, t) + 3 \, \cosh (x - 28 \, t) \right)^5} 
             \quad
\ee
which is precisely the expression for the second order ncKdV correction 
$u_2$ to the classical 2-soliton solution (\ref{KdV-2s}), obtained 
in \cite{DMH-ncKdV} in a different way.

\subsection{Further examples}
\bigskip
\noindent
{\em 1. Sine-Gordon equation.} 
Let $M = C^\infty(\mathbb{R}^2, \mathbb{C}) \otimes \Lambda_2$ and
\be
  \dl z = z_x \, \xi + {1 \over 2} (\bar{z}-z) \, \tau  \, , \qquad
   \d z = {1 \over 2} (\bar{z}-z) \, \xi + z_y \, \tau
                              \label{sG_bic}
\ee
for $z \in M^0$, where a bar indicates complex conjugation. 
Dressing $\d$ with $A = g^{-1} \td g$, where $g = e^{-i \phi/2}$ with a 
real function $\phi$, we obtain the map $ \D z = e^{i \phi/2} \, \d( e^{-i \phi/2} z) $.
Then  $(M,\dl,\D)$ is a bicomplex associated 
with the sine-Gordon equation $\phi_{xy} = \sin \phi$ \cite{DMH-bico}.
 Following scheme (A) of section 3, in order to calculate the primary 
DBT, we have to evaluate (\ref{gba}) with $g_i = e^{-i\phi_i/2}$ 
and some operator $a$. The latter has to satisfy $\td a = 0$ which 
means $\overline{a z} = a \bar{z}$ and $a_y = 0$.
 First we calculate the right side of (\ref{gba}):
\be
   g_2^{-1} \td (g_2 g_1^{-1}) \, g_1 \, z 
 = {1 \over 2} (e^{i \phi_2} - e^{i \phi_1}) \, \bar{z} \, \xi 
   - {i \over 2} (\phi_2 - \phi_1)_y \, z \, \tau  \; .
\ee
In order for this to be consistent (for all $z$) with the left 
side of (\ref{gba}), the operator $a$ must have the form 
$a z = \alpha \, \bar{z}$ with a real function $\alpha(x)$. Then 
\be
   \tdl (g_2^{-1} a \, g_1) \, z 
 = [{i \alpha \over 2} \, (\phi_1 + \phi_2)_x  + \alpha_x ] \, 
    e^{i (\phi_1+\phi_2)/2} \, \bar{z} \, \xi 
   - i \alpha \, \sin \left( {\phi_1 + \phi_2 \over 2} \right) \, z \, \tau
\ee
and (\ref{gba}) results in the following two equations,
\be
  (\phi_1 + \phi_2)_x = {2 \over \alpha} \, \sin{\phi_2 - \phi_1 \over 2} 
      + 2 i \, {\alpha_x \over \alpha}  \, , \qquad
  (\phi_2 - \phi_1)_y = 2 \alpha \, \sin{\phi_1 + \phi_2 \over 2} \; .
               \label{sG-BT}
\ee
Since we consider only {\em real} sine-Gordon solutions, we have to 
set $\alpha_x = 0$. Hence $\alpha$ has to be a real constant. 
Now we recover a famous auto-BT of the sine-Gordon equation 
\cite{Eisen09,BT,Roge+Shad82,BT-sineG}. 

Since 
\be
  R_{ij} \, z = \alpha_{ij} \, e^{i (\phi_i + \phi_j)/2} \, \bar{z} \, ,
\ee
the permutability condition (\ref{perm2A}) is satisfied with  
$\alpha_{10} = \alpha_{32} = \alpha_1$ and $\alpha_{20} = \alpha_{31} =\alpha_2$. 
The remaining permutability condition (\ref{perm1}) requires
\be
   \alpha_1 \, e^{i (\phi_0 + \phi_1)/2} 
 + \alpha_2 \, e^{i (\phi_1 + \phi_3)/2} 
 - \alpha_2 \, e^{i (\phi_0 + \phi_2)/2} 
 - \alpha_1 \, e^{i (\phi_2 + \phi_3)/2} = 0 
\ee
which is equivalent to Bianchi's `permutability theorem' for the 
sine-Gordon equation:
\be
  \phi_3 = \phi_0 + 4 \, \arctan \left[{\alpha_1 + \alpha_2 \over \alpha_1 - \alpha_2}
  \tan \left({\phi_1 - \phi_2 \over 4} \right) \right] \, .  \label{sG-perm}
\ee
This determines a solution $\phi_3$, if $\phi_1$ and $\phi_2$ are obtained from 
$\phi_0$ via the BT, i.e. the pairs $(\phi_0,\phi_1)$ and $(\phi_0,\phi_2)$ have to 
satisfy (\ref{sG-BT}).\footnote{The reader should be aware of a problem of notation 
in this section. A BT like (\ref{sG-BT}) is written in terms of solutions $\phi_1$ 
and $\phi_2$. But these are {\em not} the solutions $\phi_1$ and $\phi_2$ 
which appear in the permutability relations. See also section 1.}

\bigskip
\noindent
{\em 2. An equation related to the sine-Gordon equation.}
Let us again consider the trivial bicomplex (\ref{sG_bic}) which we used in the 
context of the sine-Gordon equation. But now we choose a different 
dressing:
\be
  \D z = \d z + [\dl,U] \, z = {1 \over 2} [(1+2 u_x) \, \bar{z} - z] \, \xi
         + [z_y + {1 \over 2} (\bar{u} - u) \, z] \, \tau 
\ee
where $U z = u \, \bar{z}$ with a field $u(x,y)$. Then we have 
$\delta^2 = 0 = \dl \D + \D \dl$ identically, while $\D^2 = 0$ is 
equivalent to
\be
   u_{xy} = {1 \over 2} (u - \bar{u}) (1 + 2u_x)  \; .   \label{dsG}
\ee
Since $\D = \d + \tdl U$, following scheme (B) we have $R = U_2 - U_1 + T$ 
with $\tdl T = 0$ for a primary DBT. The latter condition requires $T_x = 0$ 
and $\overline{Tz} = T \bar{z}$. This is satisfied with $T z = \alpha \, \bar{z}$ 
where $\alpha(y)$ is a real function. However, since the transformation 
$u_1 \mapsto u_1 + \alpha$ leaves (\ref{dsG}) invariant, we may set 
$\alpha = 0$ and obtain
\be
      R z = (u_2 - u_1) \, \bar{z} \; .
\ee
The primary DBT conditions now take the form
\be
    (\bar{u}_2 - \bar{u}_1 ) ( 1 + 2 \, u_{2x} )
 - (u_2 - u_1) ( 1 + 2 \, \bar{u}_{1x} ) &=& 0   \\
   (u_2 - u_1)_y + {1 \over 2} \, (\bar{u}_1 + \bar{u}_2 - u_1  - u_2)
   (u_2 - u_1) &=& 0  \, .
\ee
Adding the first equation to, respectively subtracting it from its complex conjugate, 
we deduce
\be
    |u_2 - u_1|_x = 0   \, ,  \qquad  
     {\bar{u}_2 - \bar{u}_1 \over u_2 - u_1} 
 =  { 1 + (\bar{u}_1 + \bar{u}_2)_x \over 1 + (u_1 + u_2)_x } \; . 
\ee
 Furthermore, one obtains $ | 1 + 2 \, u_{1x} | = | 1 + 2 \, u_{2x} |$. 

The first permutability condition (\ref{perm1}) is identically satisfied 
and the second permutability condition (\ref{perm2}) leads to
\be
    u_3 = { u_2 \, (\bar{u}_2 - \bar{u}_0) - u_1 \, (\bar{u}_1 - \bar{u}_0) 
              \over \bar{u}_2 - \bar{u}_1 }  \; . 
\ee

Comparing the operator $\D$ with the corresponding operator in the 
sine-Gordon case, we find the transformation
\be
  u_x = {1 \over 2} (e^{i \phi} - 1) \, ,  \qquad
  \phi_y = i  \, (\bar{u} - u) \, .
\ee
Eliminating $\phi$ from the above equations, we obtain (\ref{dsG}). 
If we eliminate $u$, then we obtain the sine-Gordon equation.

\bigskip
\noindent
{\em 3. Discrete sine-Gordon equation.} Let $M^0$ be the space of 
complex functions on an infinite plane square lattice and 
\be
  (\dl z)_S = (z_E - z_S) \, \xi + \kappa \, (\bar{z}_W - z_S) \, \tau \, , 
              \qquad
   (\d z)_S = \kappa \, (\bar{z}_E - z_S) \, \xi + (z_W - z_S) \, \tau            
\ee
where $\kappa$ is a real parameter. We use the notation
$z_S=z(x-1,y-1), z_E=z(x-1,y+1), z_W=z(x+1,y-1), z_N=z(x+1,y+1)$ 
(see also \cite{DMH-bico}).
Now $\d$ gets dressed with
\be
  (A z)_S = [(e^{i\phi/2} \, \td e^{-i \phi/2}) \, z]_S =
  \kappa \, e^{i(\phi_E+\phi_S)/2} \, \bar{z}_E \, \xi 
  + e^{-i(\phi_W-\phi_S)/2} \, z_W \, \tau \, .
\ee
The bicomplex condition $\tdl A=0$ then reads
\be 
    e^{i(\phi_E-\phi_N)/2} - e^{i(\phi_S-\phi_W)/2} 
 = \kappa^2 \left( e^{i(\phi_W+\phi_N)/2} - e^{i(\phi_E+\phi_S)/2} \right) 
\ee
which, multiplied by $e^{i(\phi_N-\phi_E+\phi_W-\phi_S)/4}$, produces
the discrete sine-Gordon equation \cite{disc-sine-G}
\be
   \sin [(\phi_N-\phi_E-\phi_W+\phi_S)/4] 
 = \kappa^2 \, \sin [(\phi_N+\phi_E+\phi_W+\phi_S)/4] \, .
\ee
 Following scheme (A) of section 3 with $a z = \alpha \, \bar{z}$ 
where $\alpha$ is a real constant, we find  
\be
   (R z)_S = (g_2^{-1} \, a \, g_1 \, z)_S 
 = \alpha \, e^{i (\phi_{1,S} + \phi_{2,S}) / 2} \, \bar{z}_S  \; .
\ee
Now (\ref{gba}) generates the BT
\be
     \sin {(\phi_1+\phi_2)_E - (\phi_1+\phi_2)_S \over 4} 
 &=& {\kappa \over \alpha} \, \sin{(\phi_2 - \phi_1)_E + (\phi_2-\phi_1)_S \over 4}  \\
     \sin{(\phi_2-\phi_1)_W - (\phi_2-\phi_1)_S \over 4} 
 &=& - \alpha \, \kappa \, \sin{(\phi_1+\phi_2)_W + (\phi_1+\phi_2)_S \over 4} 
\ee
(see also \cite{disc-sine-G}).

With $\alpha_{10} = \alpha_{32} = \alpha_1$ and $\alpha_{20} = \alpha_{31} =\alpha_2$, 
(\ref{perm2A}) is satisfied and the remaining permutability condition (\ref{perm2}) for 
the primary DBT reads
\be
    \alpha_1 \left( e^{i(\phi_0+\phi_1)_S/2} - e^{i(\phi_2+\phi_3)_S/2} \right)
 = \alpha_2 \left( e^{i(\phi_0+\phi_2)_S/2} - e^{i(\phi_1+\phi_3)_S/2} \right)
\ee
from which we obtain again (\ref{sG-perm}).

\bigskip
\noindent
{\em 4. Infinite Toda lattice.} Let $M^0$ be the set of real functions $z_k(t)$, 
$k \in \mathbb{Z}$, which are smooth in the variable $t$. On $M^0$ we define
\be
  (\dl z)_k = \dot{z}_k \, \tau + (z_{k+1}-z_k) \, \xi  \, , \qquad
   (\d z)_k = (z_k-z_{k-1}) \, \tau + \dot{z}_k \, \xi 
\ee
where $\dot{z} = \pa z/\pa t$. 
Together with these maps, $M = M^0 \otimes \Lambda_2$ is a trivial bicomplex 
\cite{DMH-bico}. 
Dressing $\d$ with $A = g^{-1} \d g$ where $g = e^{q_k}$, this yields a bicomplex 
for the nonlinear Toda lattice equation
\be
    \ddot{q}_k = e^{q_{k-1}-q_k} - e^{q_k-q_{k+1}}  \; .
\ee
Again, we follow scheme (A) of section 3 to determine a primary DBT. 
Let $g_1 = e^{p_k}, \, g_2 = e^{q_k}$ and $(a z)_k = \alpha \, z_{k-1}$ with a 
constant $\alpha$. Then
\be
     {1 \over \alpha} \, [\tdl (g_2^{-1} a \, g_1) z]_k 
 &=& (\dot{p}_{k-1} - \dot{q}_k) \, e^{p_{k-1}-q_k} \, 
     z_{k-1} \, \tau + (e^{p_k-q_{k+1}} - e^{p_{k-1}-q_k}) \, z_k \, \xi \\
     \left( g_2^{-1} \td (g_2 g_1^{-1}) \, g_1 \, z \right)_k 
 &=& (e^{p_{k-1}-p_k}-e^{q_{k-1}-q_k}) \, z_{k-1} \, \tau 
     + (\dot{q}_k-\dot{p}_k) \, \xi 
\ee
so that (\ref{gba}) yields
\be
  \dot{q}_k-\dot{p}_k = \alpha \, (e^{p_k-q_{k+1}}-e^{p_{k-1}-q_k})  
      \, , \quad
  \dot{p}_{k-1}-\dot{q}_k = {1 \over \alpha} \, (e^{q_k-p_k}-e^{q_{k-1}-p_{k-1}}) 
  \, .
\ee
We can absorb $\alpha$ in $p_k$ by a redefinition $p_k \mapsto p_k - \ln|\alpha|$ 
and choose the sign of $t$ such that the above equations become
\be
  \dot{q}_k - \dot{p}_k = -e^{p_k-q_{k+1}} + e^{p_{k-1}-q_k} \, , \quad
  \dot{p}_{k-1} - \dot{q}_k = -e^{q_k-p_k} + e^{q_{k-1}-p_{k-1}} \; .
                          \label{Toda_BTv1}
\ee
This is a well-known auto-BT of the Toda lattice.
 From these equations we obtain immediately
\be
  (e^{q_k-p_k})^\cdot =  -e^{q_k-q_{k+1}} + e^{p_{k-1}-p_k} \, , \quad
  (e^{p_{k-1}-q_k})^\cdot = -e^{p_{k-1}-p_k} + e^{q_{k-1}-q_k}  \, .
\ee
Adding these equations and using the Toda equation yields
\be
    (e^{q_k-p_k}+e^{p_{k-1}-q_k})^\cdot 
 = -e^{q_k-q_{k+1}}+e^{q_{k-1}-q_k} = \ddot{q}_k
\ee
and, after integration, 
\be
   \dot{q}_k = e^{q_k-p_k} + e^{p_{k-1}-q_k} - \gamma_k
\ee
with integration constants $\gamma_k$.
In a similar way, we obtain
\be
   \dot{p}_k = e^{q_k-p_k} + e^{p_k-q_{k+1}} - \tilde{\gamma}_k
\ee
with integration constants $\tilde{\gamma}_k$.
Substituting these expressions into (\ref{Toda_BTv1}), 
we find that $\gamma_k = \tilde{\gamma}_k = \gamma$ is a constant, 
and thus
\be
  \dot{q}_k = e^{q_k-p_k} + e^{p_{k-1}-q_k} - \gamma  \, , \qquad
  \dot{p}_k = e^{q_k-p_k} + e^{p_k-q_{k+1}} - \gamma
\ee
which is another form of the auto-BT of the 
infinite Toda lattice, with a parameter $\gamma$ \cite{BT-Toda}. 

In terms of $g_i = e^{q_i}$, $i=0,1,2,3$, we have
\be
    (R_{ij} z)_k = \alpha_{ij} \, e^{-q_{i,k} + q_{j,k-1}} \, z_{k-1}
\ee
and with $\alpha_{10} = \alpha_{32} = \alpha_1$ and $\alpha_{20} 
= \alpha_{31} =\alpha_2$ the permutability conditions amount to 
\be
 q_{3,k} = - q_{0,k-1} + q_{1,k} + q_{2,k} 
                 + \ln {\alpha_2 \, e^{q_{1,k-1}} - \alpha_1 \, e^{q_{2,k-1}} 
                  \over \alpha_2 \, e^{q_{1,k}} - \alpha_1 \, e^{q_{2,k}} } \; . 
\ee

\bigskip
\noindent
{\em 5. Hirota's difference equation.} For functions $z_k(u,v)$ of three discrete 
variables $k,u,v$, we set $(K z)_k(u,v) = z_{k-1}(u,v)$,  
$(U z)_k(u,v) = z_k(u+1,v)$, $(V z)_k(u,v) = z_k(u,v+1)$, and define bicomplex 
maps
\be
  \dl z = (U-1) z \, \xi + (V-1) z \, \tau \, , \qquad
   \d z = \kappa_1 \, (K U-1) z \, \xi + \kappa_2 \, (K V-1) z \, \tau \, .
\ee
With a dressing similar to that for the Toda lattice we obtain a gauge 
potential 1-form
\be
  A = e^{-q} \, \td e^q = \kappa_1 \, (e^{K U(q) - q } - 1) \, K U \, \xi 
      + \kappa_2 \, (e^{K V(q) - q } -1) \, K V \, \tau 
\ee
using the notation $K(q) = K q K^{-1}$. Now $\tdl A = 0$ with $k \mapsto k+1$ 
becomes 
\be
  \lefteqn{ \kappa_1 \left( e^{q_k(u+1,v+1)-q_{k+1}(u,v+1)} 
 - e^{q_k(u+1,v) - q_{k+1}(u,v)} \right) 
          } 
  \hspace{3cm} \nonumber \\
 &=& \kappa_2 \left( e^{q_k(u+1,v+1)-q_{k+1}(u+1,v)} 
     - e^{q_k(u,v+1) - q_{k+1}(u,v)} \right)  
\ee
which, in an equivalent form, is known as {\em Hirota's bilinear difference equation} 
\cite{Hirota-diff,Zabr97,BT-Hirota}.
 Following scheme (A), we choose $a = \alpha \, K$ with a constant $\alpha$, 
so that 
\be
   R = \alpha \, e^{-q_2} K e^{q_1} = \alpha \, e^{K(q_1)-q_2} \, K \, .
\ee
Now we obtain from (\ref{gba}) the BT
\be
  \lefteqn{ \alpha \left( e^{q_{1,k}(u+1,v) - q_{2,k+1}(u+1,v)} 
            - e^{q_{1,k}(u,v) - q_{2,k+1}(u,v)}
                   \right)  
          }
  \hspace{3cm} \nonumber \\  
  &=& \kappa_1 \left( e^{q_{2,k}(u+1,v) - q_{2,k+1}(u,v)} 
      - e^{q_{1,k+1}(u+1,v) - q_{1,k}(u,v)}
                  \right)       \\
  \lefteqn{ \alpha \left( e^{q_{1,k}(u,v+1) - q_{2,k+1}(u,v+1)} 
                          - e^{q_{1,k}(u,v)-q_{2,k+1}(u,v)}
                   \right) 
          }
  \hspace{3cm}  \nonumber  \\  
  &=& \kappa_2 \left( e^{q_{2,k}(u,v+1) - q_{2,k+1}(u,v)} 
                      - e^{q_{1,k+1}(u,v+1)-q_{1,k}(u,v)}
               \right) \, .  \quad
\ee
As a `permutability theorem', we obtain
the same formula for $q_{3,k}$ as in the Toda lattice example.

\bigskip
\noindent
{\em 6. Principal chiral model.} Let 
$M = C^\infty(\mathbb{R}^2, \mathbb{C}^m) \otimes \Lambda_2$ and
\be
   \dl z = z_t \, \tau + z_x \, \xi \, , \qquad
    \d z = z_t \, \tau - z_x \, \xi 
\ee
for $z \in M^0$. Let $G$ be a group of $m \times m$ matrices. 
Dressing $\d$ with the gauge potential 1-form $A = g^{-1} \td g$, 
where $g \in G$, this yields a bicomplex for the principal 
chiral model field equation  
\be
   (g^{-1}g_x)_t + (g^{-1}g_t)_x = 0   \label{pcm_eq}
\ee
which is $\tdl A = 0$ (see also \cite{DMH-bico}). 
Hence, we follow scheme (A). With
\be
     g_2^{-1} \td(g_2 g_1^{-1}) \, g_1
 &=& ( g_2^{-1} g_{2t} - g_1^{-1} g_{1t} ) \, \tau
     - ( g_2^{-1} g_{2x} - g_1^{-1} g_{1x} ) \, \xi \\
     \dl (g_2^{-1} a g_1)
 &=& (g_2^{-1} a g_1)_t \, \tau 
     + (g_2^{-1} a g_1)_x \, \xi \, ,
\ee
the primary DBT condition (\ref{gba}) and $\td a = 0$ requires 
$a$ to be a constant matrix.  
 Now (\ref{gba}) results in the following two equations,
\be
  g_2^{-1} g_{2t} - g_1^{-1} g_{1t} = (g_2^{-1} a \, g_1)_t 
               \, , \qquad
  g_2^{-1} g_{2x} - g_1^{-1} g_{1x} = - (g_2^{-1} a \, g_1)_x 
               \, .
\ee
If $a$ is invertible, the transformation $g_2 \mapsto a \, g_2$ 
leaves (\ref{pcm_eq}) invariant and eliminates $a$ from the 
last equations. This is no longer possible if $g$ is 
constrained to some subgroup $G \subset GL(m,\mathbb{C})$. 
 For $G = U(m)$, the last equations read
\be
    g_2^\dagger \, g_{2t} - g_1^\dagger \, g_{1t} 
  = (g_2^\dagger a \, g_1)_t \, , \qquad
    g_2^\dagger \, g_{2x} - g_1^\dagger \, g_{1x} 
  = - (g_2^\dagger a \, g_1)_x  \; . \label{pcm_uni_BT}
\ee
Adding the Hermitian conjugates, the resulting equations 
can be integrated. Using $g_i^\dagger \, g_i = I$, they lead to
\be 
  g_2^\dagger \, a \, g_1 + g_1^\dagger \, a^\dagger \, g_2 
   = C  \label{pcm_uni_constr}
\ee
with a constant real matrix $C$. Assuming again that $a$ is invertible, 
it can be written as a product $u \, h$ of a unitary matrix $u$ with 
a Hermitian matrix $h$. A redefinition $g_2 \mapsto u \, g_2$ 
(which leaves the field equations and the unitarity constraint invariant) 
then eliminates $u$ from the above equations. Hence we can assume that 
$a$ is Hermitian. Then we find $[ g_2^\dagger \, a \, g_1 , C ] = 0$ 
(for all $g_1, g_2$) and thus $C = c \, I$ with $c \in \mathbb{R}$. 
 For $a = \alpha \, I$ with $\alpha \in \mathbb{R}$, we now recover from 
(\ref{pcm_uni_BT}) and (\ref{pcm_uni_constr}) a well-known auto-BT 
for the unitary principal chiral model \cite{BT-pcm,Cher96}. 

With $R_{ij} = g_i^{-1} \, a_{ij} \, g_j$, the permutability conditions take 
the following form,
\be
    g_3 = (a_{31} \, g_1 - a_{32} \, g_2) \, g_0^{-1} \, (g_2^{-1} \, a_{20} 
              - g_1^{-1} \, a_{10} )^{-1} \, , \qquad
    a_{31} \, a_{10} = a_{32} \, a_{20} \; .
\ee

\bigskip
\noindent
{\em 7. Nonlinear Schr\"odinger equation.} Let 
$M = C^\infty(\mathbb{R}^2, \mathbb{C}^2) \otimes \Lambda_2$ with
\be
  \dl z = z_x \, \tau + {i \over 2} (\sigma_3-I) \, z \, \xi 
             \, , \qquad
  \d z  = z_t \, \tau + z_x \, \xi 
\ee
for $z \in M^0$. Furthermore, let $A = - V \, \tau - U \, \xi$ with
\be
  U = \left(\begin{array}{cc} 0 & -\bar{\psi} \\ \psi & 0 \end{array}\right) \, ,
        \qquad
  V = i \, (U_x + U^2) \, \sigma_3 =
        i \left( \begin{array}{cc} -|\psi|^2 & \bar{\psi}_x \\ 
                 \psi_x & |\psi|^2
                 \end{array} \right) \, .
\ee
This dressing for $\d$ yields a bicomplex for the nonlinear Schr\"odinger 
equation
\be
   i \psi_t + \psi_{xx} + 2 |\psi|^2 \psi = 0 
\ee
(see also \cite{DMH-bico}). 
It is helpful to note that $\sigma_3 \, U + U \, \sigma_3 = 0$ and 
$U^\dagger =-U$.
The primary DBT conditions (\ref{primaryDBT}) now imply the following 
four equations
\be
   R_x &=& V_1 - V_2 \, ,      \label{ns1}   \\
   {i \over 2} \, [\sigma_3 , R] &=& U_1 - U_2 \, ,  \label{ns2} \\
   R_t &=& V_2 \, R - R \, V_1 \, , \label{ns3} \\
   R_x &=& U_2 \, R - R \, U_1    \label{ns4}
\ee
where $U_i, V_i$ are $U,V$ with $\psi$ replaced by $\psi_i$, $i=1,2$. 
Decomposing $R = R^+ + R^-$ such that 
$\sigma_3 \, R^\pm = \pm R^\pm \, \sigma_3$, we obtain 
$R^+ = k \, I +  i \, r \, \sigma_3$ with functions $k$ and $r$. 
 Furthermore, (\ref{ns2}) implies
\be
   R^- = i \, (U_1 - U_2) \, \sigma_3  \, .
\ee
 From (\ref{ns1}) we obtain $k_x = 0$ and
\be
  R^+_x = i \, (|\psi_2|^2 - |\psi_1|^2) \, \sigma_3 \, .
\ee
Hence
\be
    R = \tilde{R} + k \, I   \, , \quad
    \tilde{R} = i \left( \begin{array}{cc} r & \bar{\psi}_1 - \bar{\psi}_2 \\
     \psi_1 - \psi_2 & -r \end{array} \right)
\ee
with
\be
   r_x = |\psi_2|^2 - |\psi_1|^2 \, .   \label{r1}
\ee
 From (\ref{ns4}) we obtain $\bar{k} = k$, $\bar{r}=r$ and
\be
  (\psi_1 - \psi_2)_x = r \, (\psi_1 + \psi_2) + i \, k \, (\psi_1 - \psi_2) \, .   
                                                        \label{r2}
\ee
With the help of (\ref{r1}), this leads to 
\be
    (|\psi_2-\psi_1|^2)_x = 2 r \, (|\psi_1|^2-|\psi_2|^2) 
                                     = -2 r r_x
\ee
and, after integration, 
\be
  r = \pm \sqrt{\alpha^2 - |\psi_2-\psi_1|^2} 
\ee
with an integration `constant' $\alpha^2(t)$. As a consequence, 
(\ref{r2}) becomes
\be
  (\psi_1 - \psi_2)_x = \pm (\psi_1 + \psi_2) \sqrt{\alpha^2 - |\psi_2-\psi_1|^2}
                                   + i \, k \, (\psi_1 - \psi_2)   \; .       \label{bns1}
\ee
 From (\ref{ns3}) we obtain $k_t = 0$, so that $k$ has to be a constant, and
\be
  (\psi_1 - \psi_2)_t &=& \pm i \, (\psi_1 + \psi_2)_x \sqrt{\alpha^2 - |\psi_2-\psi_1|^2}
    + {i \over 2} (\psi_1 - \psi_2) \, (|\psi_1 + \psi_2|^2 + |\psi_2 - \psi_1|^2)
           \nonumber \\
  & &   - k \, (\psi_1 - \psi_2)_x \, .   \label{bns2}
\ee
The last two equations constitute a well-known auto-BT of the 
nonlinear Schr\"odinger equation 
\cite{Lamb74,Roge+Shad82,BT-NLS,BLP81,Grec84}.\footnote{The terms 
proportional to the parameter $k$ are due to the symmetry transformation 
$\psi(t,x) = e^{i \, (-k^2 t +k x)} \, \hat{\psi}(t,x-2k t)$ 
of the NLS equation. See also \cite{Roge+Shad82}, p. 68.}

Let us turn to the permutability conditions. 
 First we note that $\tilde{R}_{ij}$ is anti-Hermitian, traceless and satisfies
\be
     \tilde{R}_{ij} \, \tilde{R}_{kl} + \tilde{R}_{kl} \, \tilde{R}_{ij} 
  = [ (r_{ij} - r_{kl})^2 + | \psi_i - \psi_j - \psi_k + \psi_l |^2 
     - \alpha_{ij}^2 - \alpha_{kl}^2 ] \, I  \, .  \label{NLS_RR}
\ee
In particular, we have $\tilde{R}_{ij}^2 = - \alpha_{ij}^2 \, I$. 
Using $\mbox{trace}(\tilde{R}_{ij}) = 0$, (\ref{perm1}) splits 
into
\be
     \tilde{R}_{32} &=& \tilde{R}_{31} + \tilde{R}_{10} - \tilde{R}_{20} 
                                      \label{Rt32}   \\
     k_{32} &=& k_{31} + k_{10} - k_{20} \; .  \label{k32}
\ee
(\ref{perm2}) reads
\be
    \tilde{R}_{31} \, \tilde{R}_{10} - \tilde{R}_{32} \, \tilde{R}_{20}
   + k_{31} \, \tilde{R}_{10} + k_{10} \, \tilde{R}_{31}
   - k_{32} \, \tilde{R}_{20} - k_{20} \, \tilde{R}_{32}
   + k_{31} \, k_{10} - k_{32} \, k_{20} = 0 
\ee
and, using (\ref{Rt32}), 
\be
    \tilde{R}_{31} (\tilde{R}_{10} - \tilde{R}_{20} + k_{10} - k_{20})
 &=&  \tilde{R}_{10} \, \tilde{R}_{20} + (k_{20} - k_{31}) \, \tilde{R}_{10} 
           + (k_{32} - k_{20}) \, \tilde{R}_{20}           \nonumber \\
  & & + (\alpha_{20}^2 + k_{32} \, k_{20} - k_{31} \, k_{10}) \, I  \; .
\ee
With the help of 
\be
     (\tilde{R}_{10} - \tilde{R}_{20} + k_{10} - k_{20})^{-1} 
 = - (\tilde{R}_{10} - \tilde{R}_{20} - k_{10} + k_{20})/ \rho
\ee
where
\be
     \rho =  | \psi_1 - \psi_2|^2 + (r_{10} - r_{20})^2 + (k_{10} - k_{20})^2 
\ee
this leads to
\be
  \tilde{R}_{31} = {1 \over \rho} \left( a \, I + b \, \tilde{R}_{10} 
     + c \, \tilde{R}_{20} - 2 (k_{20} - k_{10}) \, \tilde{R}_{10} \, \tilde{R}_{20} 
      \right)                      \label{Rt31}
\ee
with
\be
      a &=& (k_{20} - k_{10}) [ | \psi_1 - \psi_2|^2 + (r_{10} - r_{20})^2 ]
                  + (k_{32} - k_{31}) \, \alpha_{10}^2          \nonumber \\
         & &  - (k_{20} - k_{10}) (k_{32} \, k_{20} - k_{31} \, k_{10} 
                  + \alpha_{20}^2)  \\
      b &=& - | \psi_1 - \psi_2|^2 - (r_{10} - r_{20})^2 + \alpha_{10}^2 - \alpha_{20}^2
                  + (k_{20} - k_{10}) (k_{20} - k_{31})    \nonumber \\
         & &  - k_{32} \, k_{20} + k_{31} \, k_{10}  \\
      c &=&  \alpha_{20}^2 - \alpha_{10}^2
                  + (k_{20} - k_{10}) (k_{20} - k_{32}) 
                  + k_{32} \, k_{20} - k_{31} \, k_{10}    \; .
\ee
Together with $\tilde{R}_{31}^2 = - \alpha_{31}^2 \, I$ this implies
\be
     k_{32} = k_{10}
\ee
and thus $k_{31} = k_{20}$ by use of (\ref{k32}), and moreover
\be
     \alpha_{31}^2 = \alpha_{20}^2 \; .
\ee
Now one can derive from (\ref{Rt31}) the following superposition formula: 
\be
           \psi_3 
  &=& \psi_0 + {1 \over \rho} \big\{ 
           [ \alpha_{20}^2 - \alpha_{10}^2 + (k_{20} - k_{10}) ^2 
            + 2 i \, (k_{20} - k_{10}) \, r_{10} ] (\psi_2 - \psi_0)      \nonumber \\
   & &  + [ \alpha_{10}^2 - \alpha_{20}^2 + (k_{10} - k_{20}) ^2 
            + 2 i \, (k_{10} - k_{20}) \, r_{20} ] (\psi_1 - \psi_0)  \big\}  \, .
\ee
The remaining equations also require $\alpha_{32}^2 + \alpha_{20}^2 
= \alpha_{31}^2 + \alpha_{10}^2$ and thus $\alpha_{32}^2 = \alpha_{10}^2$. 
With the relations between the parameters of the auto-BT derived above, we 
recover the permutability theorem as formulated in \cite{Grec84}.

Starting with the trivial NLS solution $\psi_0 = 0$, the BT  (\ref{bns1}), (\ref{bns2}) 
determines the one-soliton solution 
\be
      \psi(x,t) = \alpha \, e^{i \, [ k \, x + (\alpha^2 - k^2) \, t + \varphi ]} / 
                     \mbox{cosh}[ \alpha \, (x-x_0) - 2 \, \alpha \, k \, t ] 
\ee
with real constants $\varphi$ and $x_0$. Then 
$r = \alpha \, \mbox{tanh}[ \alpha \, (x-x_0) - 2 \, \alpha \, k \, t ] $.
Let $\psi_j$, $j=1,2$, be two such solutions with 
parameters $\alpha_j = \alpha_{j0}, k_j = k_{j0}$, $\varphi_j$ and $x_j$ 
(replacing $x_0$). Then the above formula for $\psi_3$ determines a two-soliton 
solution of the NLS equation. 

\bigskip
\noindent
{\em 8. Discrete nonlinear Schr\"odinger equation.}
Let $M^0$ be the set of $\mathbb{C}^2$-valued functions $z_k(t)$, 
$k \in \mathbb{Z}$, which are smooth in the time variable $t$. 
On $M^0$ we define
\be
 (\dl z)_k = (z_{k+1}-z_k) \, \tau 
             - {i \over 2} (\sigma_3-I) \, z_k \, \xi 
			 \, ,    \qquad
 (\d z)_k = \dot{z}_k \, \tau + (z_k-z_{k-1}) \, \xi 
\ee
where a dot denotes a time derivative. This determines a trivial 
bicomplex. Now we dress $\d$ to
\be
(\D z)_k = (\dot{z}_k - V_k z_k) \, \tau 
          + (z_k - z_{k-1} - U_k z_{k-1}) \, \xi \, .
\ee
The bicomplex conditions for $\dl$ and $\D$ are then equivalent to
\be
  U_{k+1} - U_k - {i \over2} \, [\sigma_3 , V_k] &=& 0
                  \label{dnls_bc1}    \\
  \dot{U}_k - (V_k-V_{k-1}) + U_k V_{k-1} - V_k U_k &=& 0 \, .  
                  \label{dnls_bc2}
\ee
With the decomposition $V_k = V_k^+ + V_k^-$ such that 
$\sigma_3 V_k^\pm = \pm V_k^\pm \sigma_3$, (\ref{dnls_bc1}) implies
\be
  V_k^- = i \, (U_{k+1}-U_k) \, \sigma_3 \, .
\ee
In the following we assume that $\sigma_3 \, U_k = - U_k \, \sigma_3$. 
Now we decompose (\ref{dnls_bc2}) into $\pm$ parts and find
\be
  V_k^+ = i \, U_{k+1} U_k \, \sigma_3  \, ,  \quad
  i \, \dot{U}_k \, \sigma_3 + (U_{k+1} + U_{k-1} - 2U_k)
  - (U_k{}^2 \, U_{k-1} + U_{k+1} \, U_k{}^2) = 0  \, .   
                    \label{dnls_U}
\ee
Assuming furthermore $U_k{}^\dagger = - U_k$, we can write
\be
  U_k = \left(\begin{array}{cc} 0 & -\bar{\psi}_k \\ 
                           \psi_k & 0 
			 \end{array} \right)
\ee
so that
\be
  V_k = i \left( \begin{array}{cc} 
       - \bar{\psi}_{k+1}\psi_k & \bar{\psi}_{k+1}-\bar{\psi}_k \\
              \psi_{k+1}-\psi_k & \psi_{k+1}\bar{\psi}_k 
	            \end{array} \right)
\ee
and from the second of equations (\ref{dnls_U}) we obtain
\be
  i \, \dot{\psi}_k + (\psi_{k+1} + \psi_{k-1}-2\psi_k) 
  - |\psi_k|^2 (\psi_{k+1} + \psi_{k-1}) = 0 
\ee
which is the {\em discrete nonlinear Schr\"odinger equation} of Ablowitz and 
Ladik \cite{Ablo+Ladi76}.
The equations (\ref{primaryDBT}), which determine a primary DBT, 
with $(R z)_k = P_k z_{k-1}$ lead to 
\be
  P_{k+1}-P_k &=& - V_{2,k} + V_{1,k}   \label{dnls_BT1}  \\
  {i \over 2} \, [\sigma_3,P_k] &=& -U_{2,k} + U_{1,k}   \label{dnls_BT2}\\
  \dot{P}_k &=& V_{2,k} \, P_k - P_k \, V_{1,k-1}  \label{dnls_BT3}\\
  P_k-P_{k-1} &=& U_{2,k} \, P_{k-1} - P_k \, U_{1,k-1}  \, .  \label{dnls_BT4}
\ee
Decomposing $P_k = P_k^+ + P_k^-$, (\ref{dnls_BT2}) implies  
\be
  P_k^- = i \, (U_{1,k} - U_{2,k}) \, \sigma_3  
\ee
and from (\ref{dnls_BT1}) we obtain
\be
  P_{k+1}^+ - P_k^+ = i \, (U_{1,k+1} \, U_{1,k} - U_{2,k+1} \, U_{2,k}) \, \sigma_3 
      \, . \label{dnls_Pk}
\ee
Setting 
\be 
 P_k^+ = i \left(\begin{array}{cc} p_k & 0  \\
                                     0 & \bar{p}_k
                  \end{array} \right) \, \sigma_3 
\ee
we have
\be
 P_k = i \, \left(\begin{array}{cc} 
                           p_k & \bar{\psi}_{1,k} - \bar{\psi}_{2,k} \\
       \psi_{1,k} - \psi_{2,k} & - \bar{p}_k 
	              \end{array} \right)
\ee
and (\ref{dnls_Pk}) becomes
\be
 p_{k+1}-p_k = \bar{\psi}_{2,k+1} \, \psi_{2,k} - \bar{\psi}_{1,k+1} \, \psi_{1,k} 
       \, .   \label{dnls_p1}
\ee
Now we obtain from (\ref{dnls_BT4}) 
\be
   (\psi_1 - \psi_2)_k - (\psi_1 - \psi_2)_{k-1} 
 = p_{k-1} \, \psi_{2,k} + \bar{p}_k \, \psi_{1,k-1}  \label{dnls_BTa}
\ee
and (\ref{dnls_BT3}) leads to
\be
   (\psi_{1,k} - \psi_{2,k})\dot{ } &=&  i \, (\psi_{1,k} - \psi_{2,k}) ( \bar{\psi}_{1,k} \psi_{1,k-1} 
  + \bar{\psi}_{2,k} \psi_{2,k+1})  \nonumber  \\
 & & + i \, p_k \, (\psi_{2,k+1} - \psi_{2,k}) 
         + i \, \bar{p}_k \, (\psi_{1,k} - \psi_{1,k-1})  \, .  \label{dnls_BTb}  
\ee
In order to obtain a BT one has to eliminate $p_k$ from (\ref{dnls_BTa}) and 
(\ref{dnls_BTb}) with the help of (\ref{dnls_p1}). However, there seems to be no convenient 
way to achieve that.

\bigskip
\noindent
{\em 9. A generalized Volterra equation.}
Let $M^0$ be the set of functions $z_n(t)$, $n \in \mathbb{Z}$, which are smooth in 
the variable $t$ und which have values in $\mathbb{C}^m$, $m \in \mathbb{N}$. 
On $M^0$ we define 
\be
  (\dl z)_n = (z_n-z_{n-k}) \, \xi + \dot{z}_n \, \tau   \, , \quad
  (\d z)_n = (z_{n+1}-z_n) \, \xi + (z_{n+k+1}-z_n) \, \tau
\ee
for some fixed  $k \in \mathbb{Z}$. Then $M = M^0 \otimes \Lambda_2$ 
together with $\delta$ and $\d$ is a trivial bicomplex. Now we introduce a dressing:
\be
  (\D z)_n = (g^{-1} \, \d \, g z)_n =  (g_n^{-1}g_{n+1}z_{n+1}-z_n) \, \xi
 + (g_n^{-1}g_{n+k+1}z_{n+k+1}-z_n) \, \tau  
\ee
with an invertible $m \times m$ matrix $g$, depending on $t$ and the discrete 
variable $n$. Introducing the abbreviation
\be
     V_n = g_n^{-1} \, g_{n+1} \, ,     \label{gVolt_V_g}
\ee
the operator $\D$ can be expressed as follows,
\be
   (\D z)_n = (V_n \, z_{n+1} - z_n) \, \xi + (V_n V_{n+1} \cdots
                   V_{n+k}\, z_{n+k+1} - z_n) \, \tau \, .
\ee
The only nontrivial bicomplex condition is $\dl \D + \D \dl = 0$. It results in 
the generalized Volterra equation
\be
  \dot{V}_n = V_n V_{n+1} \cdots V_{n+k} - V_{n-k}V_{n-k+1} \cdots V_n 
                      \label{gVolt}
\ee
which is also known as one of the Bogoyavlenski\u{i} lattices (\cite{Bogo88}, 
see also \cite{Suri97}). For $k=1$ it reduces to (a matrix version of) the 
Volterra equation 
\be
    \dot{V}_n = V_n \, V_{n+1} - V_{n-1} \, V_n \, .
\ee

In order to elaborate primary DBTs, we have to solve the equations 
(\ref{primaryDBT}). 
With the ansatz $(R z)_n = r_n \, z_{n+k+1}$, we obtain the equations
\be
 r_n-r_{n-k} = V_{2,n} - V_{1,n} \, , \qquad
     \dot{r}_n = V_{2,n} \cdots V_{2,n+k} - V_{1,n} \cdots V_{1,n+k}
                                        \label{gVolt_r}
\ee
and
\be
   V_{2,n} \, r_{n+1} = r_n \, V_{1,n+k+1} \, .
\ee
Expressing $V$ back in terms of $g$, we can `integrate' the last
equation and obtain
\be
    r_n = g_{2,n}^{-1} \, a \, g_{1,n+k+1} 
\ee
with an arbitrary $m \times m$ matrix $a(t)$. Inserted in the two equations 
(\ref{gVolt_r}), this leads to the following BT for the generalized Volterra equation:
\be
     g_{2,n}^{-1}\, g_{2,n+1} - g_{1,n}^{-1} \, g_{1,n+1} 
 &=& g_{2,n}^{-1} \, a \, g_{1,n+k+1} - g_{2,n-k}^{-1} \, a \, g_{1,n+1}
                         \label{gVolt_BT1}   \\
     {d \over dt} (g_{2,n}^{-1} \, a \, g_{1,n+k+1})
 &=& g_{2,n}^{-1} \, g_{2,n+k+1} - g_{1,n}^{-1} \, g_{1,n+k+1} \, .
                         \label{gVolt_BT2}
\ee

As a permutability relation, we obtain 
\be
  g_{3,n} = (a_2 \, g_{1,n+k+1} - a_1 \, g_{2,n+k+1}) \, g_{0,n+k+1}^{-1}
                  \, (g_{2,n}^{-1} \, a_2 - g_{1,n}^{-1} \, a_1)^{-1}
\ee
where $a_{10} = a_{32} = a_1$ and $a_{20} = a_{31} = a_2$ and $[a_1,a_2]=0$.
\vskip.1cm

Let us now restrict our considerations to the {\em scalar} case where $m=1$. 
Inserting
\be
   g_n = {f_n \over f_{n-k-1}}    \label{gVolt_gf}
\ee
into the BT part (\ref{gVolt_BT1}) yields
\be
  { f_{1,n} \, f_{2,n+1} - a \, f_{1,n+k+1} \, f_{2,n-k} \over f_{1,n+1} \, f_{2,n}} 
 = {f_{1,n-k-1} \, f_{2,n-k} - a \, f_{1,n} \, f_{2,n-2k-1} \over f_{1,n-k} \, f_{2,n-k-1}}
\ee
which can be `integrated':
\be
     f_{1,n} \, f_{2,n+1} - a \, f_{1,n+k+1} \, f_{2,n-k} 
 = \beta_i \, f_{1,n+1} \, f_{2,n}
\ee
where $\beta_i$ with $n = i \, \mbox{mod}(k+1)$ are `constants of
integration'. The complementary BT part (\ref{gVolt_BT2}) with $\dot{a}=0$ 
becomes
\be
  { a \, D_t ( f_{1,n+k+1} \cdot f_{2,n} )
  - f_{1,n} \, f_{2,n+k+1} \over f_{1,n+k+1} \, f_{2,n}}
 =  { a \, D_t ( f_{1,n} \cdot f_{2,n-k-1} )
  - f_{1,n-k-1} \, f_{2,n} \over f_{1,n} \, f_{2,n-k-1}}
\ee
using Hirota's bilinear operator $D_t (f \cdot h) = \dot{f}h - f \dot{h}$. 
This can also be integrated with the result
\be
 a \, D_t ( f_{1,n+k+1} \cdot f_{2,n} ) - f_{1,n} \, f_{2,n+k+1} 
 = \gamma_i \, f_{1,n+k+1} \, f_{2,n}
\ee
where $\gamma_i$ with $n=i \, \mbox{mod}(k+1)$ are `constants of
integration'. Here we have obtained a BT in Hirota's bilinear form.

\section{Harry Dym equation and equivalence transformations of bicomplexes}
\setcounter{equation}{0}
Let $M = C^\infty(\mathbb{R}^2, \mathbb{R}) \otimes \Lambda_2$. With
\be
 \Dl z &=& [ \varphi \, z_x - {1 \over 2} \varphi_x \, z ] \, \xi 
    + [ 3 \varphi^2 \, z_{xx} + {3 \over 4} (\varphi_x{}^2 
        - 2 \varphi \, \varphi_{xx}) \, z ] \, \tau \, ,  \\
 \D z &=& \varphi^2 \, z_{xx} \, \xi + (z_t + 4 \varphi^3 \, z_{xxx} 
     + 6 \varphi^2 \, \varphi_x \, z_{xx}) \, \tau
\ee
we obtain $\Dl^2 = 0$ identically, and 
\be
  \D^2 &=& \xi \, \tau \, \{- 2 \, \varphi \, [ \varphi_t + \varphi^3 \, \varphi_{xxx} ] 
           \, \pa_x^2 \} \\
          \Dl \, \D + \D \, \Dl 
 &=& \xi \, \tau \, \{ {1 \over 2} \, [ \varphi_t + \varphi^3 \, \varphi_{xxx} ]_x 
          - [ \varphi_t + \varphi^3 \, \varphi_{xxx} ] \, \pa_x \}
\ee
so that the bicomplex conditions are equivalent to the Harry Dym (HD) equation
\be
   \varphi_t + \varphi^3 \, \varphi_{xxx} = 0 \; .
\ee

A relation between the HD and the KdV equation has been the subject of 
several publications \cite{HBC89,HD-KdV}. In the following, we show how such a 
relation emerges in our bicomplex framework. This is an instructive example 
of the application of equivalence transformations to bicomplexes. 
With the gauge transformation 
\be  
    \Dl' = \varphi^{-1/2} \Dl \, \varphi^{1/2} \, , \qquad
     \D' = \varphi^{-1/2} \D \, \varphi^{1/2}
\ee 
(assuming $\varphi$ to vanish nowhere) we obtain an equivalent bicomplex with
\be
  \Dl' z &=& (\varphi \pa_x) z \, \xi + 3 (\varphi \pa_x)^2 z \, \tau   \\
   \D' z &=& [ (\varphi \pa_x)^2 + {1 \over 4} (2 \varphi \varphi_{xx} - \varphi_x{}^2) ]
                     \, z \, \xi   \nonumber \\
           & & + [ \pa_t + 4 (\varphi \pa_x)^3 
                   + (\varphi \pa_x) (2 \varphi \varphi_{xx} - \varphi_x{}^2)
                   + {1 \over 2} \varphi^{-1} \, \varphi_t ] \, z \, \tau  \, .
\ee
Next we perform a change of coordinates $x = v(s,y), \, t = s$ such that 
$v_y = \varphi$. As a consequence, $\pa_y = v_y \, \pa_x = \varphi \, \pa_x$, 
$\pa_t = \pa_s + y_t \, \pa_y$, and 
$0 = x_t = v_s + y_t \, v_y$ so that $y_t = - v_s/v_y$.  
Writing $z(t,x) = \zeta(s,y)$, the bicomplex maps $\Dl'$ and $\D'$ take the 
following form in the new coordinates,
\be
  \Dl' \zeta = \zeta_y \, \xi + 3 \, \zeta_{yy} \, \tau \, ,  \quad
  \D' \zeta = (\zeta_{yy}- u \, \zeta) \, \xi 
     + (\zeta_s + 4 \, \zeta_{yyy} - p \, \zeta_y - q \, \zeta) \, \tau  
                  \label{HD_bcmaps'}
\ee
where
\be
   p = 4 \, u + {v_s \over v_y}  \, ,  \quad
   q = ({1 \over 2} \, p + 6 \, u )_y  \, ,  \quad
    u =  - {1 \over 2} {v_{yyy} \over v_y}
           + {3 \over 4} \left({v_{yy} \over v_y} \right)^2  
       = - {1 \over 2} \, {\cal S}_y v                    \label{HD_pqu}
\ee
and ${\cal S}_y$ denotes the Schwarzian derivative. The bicomplex 
conditions for the maps (\ref{HD_bcmaps'}) reduce to
\be
     p = 6 \, u          \label{HD_p}
\ee
where $u$ has to satisfy the KdV equation
\be
   u_s + u_{yyy} - 6 \, u \, u_y = 0   \, .    
\ee
(\ref{HD_p})  together with (\ref{HD_pqu}) yields\footnote{This is 
the simplest case of a Krichever-Novikov equation \cite{KNeq}.} 
\be
  v_s = - ({\cal S}_y v) \, v_y      \label{dh}
\ee
which must be equivalent to the HD equation. The KdV equation 
for $u$ is satisfied as a consequence of this equation. 
\vskip.1cm

In terms of $\psi = 1/\sqrt{v_y}$ the definition of $u$ reads $\psi_{yy} = u \, \psi$.
Given a HD solution $\varphi(t,x)$, we have to invert $y(t,x) = \int (1/\varphi) \, dx$ 
to determine $x = v(s,y)$. Then $u =  \psi_{yy}/ \psi$ is a KdV solution. 
Now we can use a KdV-BT to construct a new solution
$\hat{u}$ of the KdV equation. After solving $\hat{\psi}_{yy} = \hat{u} \, \hat{\psi}$ 
for $\hat{\psi}$, we have to invert $\hat{v} = \int (1 /\hat{\psi}^2) \, d y$ to find 
$\hat{y} = y(t,x)$. Then $\hat{\varphi} = 1/\hat{y}_x$ is again a solution of the HD equation.
In particular, if $\chi$ satisfies the first of equations (\ref{KdV_BT_chi}), i.e., 
$\chi_{yy} = (u - \alpha) \, \chi$ with a function $\alpha(s)$, 
then $\hat{\psi} = \psi_y - (\ln \chi)_y \psi$ (which is a Darboux transformation 
\cite{Matv+Sall91}) satisfies $\hat{\psi}_{yy} = \hat{u} \, \hat{\psi}$ 
with $\hat{u} = u - 2 (\ln \chi)_{yy}$. 
If $\chi$ also satisfies the second of equations (\ref{KdV_BT_chi}) (with $t,x$ 
replaced by $s,y$), then $\hat{u}$ is a KdV solution and from $\hat{\psi}$ 
we obtain a HD solution.
\vskip.1cm

As an example, let us start with the trivial solution $\varphi = 1$ of the HD equation. 
Then we obtain $y = x+a(t)$ and thus $v = y-a(s)$ with an `integration constant' $a$. 
Then we have $\psi = 1$ and consequently $u = 0$ which trivially solves the KdV equation. 
 Furthermore, the equation $\chi_{yy} = (u+k^2) \chi$ with a constant $k$ and also the second of 
equations (\ref{KdV_BT_chi})  is solved by $\chi = \chi_0 \, \cosh(k y-4k^3 s)$. 
Now we find $\hat{\psi} = - k \tanh(k y - 4k^3s)$ and $\hat{v} = y/k^2 - \coth (k y - 4 k^3s)/k^3$. 
$\hat{u} = -2 k^2 \, \mbox{sech}^2(k y - 4k^3 s)$ is the 1-soliton KdV solution. 
In order to obtain a HD solution, we have to solve $x = \hat{v}(s,y)$ 
for $y$, which results in a function $\hat{y}(t,x)$ with $t=s$. This cannot be 
done explicitly, but we find $\hat{\varphi} = k^{-2} [1 + 1/\sinh^2 (4k^3t - k \hat{y}(t,x))]$
which indeed solves the HD equation.

\vskip.1cm
Of course, one can try to solve the auto-DBT condition for the bicomplex 
$(M,\Dl,\D)$ associated with the HD equation, using (\ref{Q_Nth}). 
This turns out to be rather difficult since already the solution 
for $Q^{(0)}$ is a non-polynomial differential operator. 
It appears to be more convenient to work with the equivalent bicomplex 
$(M,\Dl',\D')$. However, the latter is tied to a less convenient form of the 
HD equation.

\section{Conclusions}
\setcounter{equation}{0}
We have introduced the concept of a Darboux-B\"acklund transformation 
(DBT) of a bicomplex and demonstrated in several examples how 
BTs for integrable models are easily obtained using this simple and 
universal construction. Once a bicomplex formulation 
is found for some equation, it is straightforward, in general, 
to apply this method. The bicomplex structure does not guarantee 
a `decent' BT, however. In some cases, the resulting correspondence 
between solutions appears to be practically not of much help (cf 
the example of the discrete NLS equation in section 4.2). 
\vskip.1cm

Higher than primary DBTs have not been sufficiently elaborated in this 
work, with the exception of the Liouville example in section 2. 
In the KdV case, the secondary DBT turned out to be a composition 
of two primary DBTs. More precisely, for one choice of sign of a 
real parameter this can only be achieved if one generalizes the 
primary DBTs to include complex transformations.
Although the latter do not, in general, generate real solutions from 
real solutions, their composition does. Hence, if we reduce our framework 
to real solutions and real maps, {\em not} all of the secondary DBTs are 
compositions of primary DBTs. Corresponding results certainly also 
hold for higher than secondary DBTs in the KdV case. This shows that, 
in general, we should not expect compositions of primary DBTs 
to exhaust the hierarchy of DBTs.
\vskip.1cm

Suppose we have three equations $EQ_i$, $i=1,2,3$, which are reductions 
of equations $\widehat{EQ}_i$ with bicomplex formulations. 
If $u_i$ is a solution of $EQ_i$, then $u_i$ is also a solution 
of $\widehat{EQ}_i$.  
Let ${\cal S}_i$ and $\hat{\cal S}_i$ denote the solution spaces of $EQ_i$ 
and $\widehat{EQ}_i$, respectively. Suppose there are BTs 
$\widehat{BT}_{21} : \hat{\cal S}_1 \rightarrow \hat{\cal S}_2$ and 
$\widehat{BT}_{32} : \hat{\cal S}_2 \rightarrow \hat{\cal S}_3$ 
(see Fig.~2) which are determined by primary DBTs.
Let $u_1 \in {\cal S}_1$ and thus also $u_1 \in \hat{\cal S}_1$. 
Applying $\widehat{BT}_{32} \widehat{BT}_{21}$ to it, yields some 
$\hat{u}_3 \in \hat{\cal S}_3$ which can be projected to some 
$u_3 \in {\cal S}_3$. 
Not all of such composed and then reduced maps will be trivial and 
not all of them should be expected to be obtainable from primary 
DBTs of the reduced bicomplexes.
However, such maps may be recovered as higher than primary DBTs 
between the initial and the final reduced bicomplex.
A very simple example is indeed provided by the KdV equation 
mentioned above. In this case, all three equations $EQ_i$ are given by 
the {\em real} KdV equation and $\widehat{EQ}_i$ is the complex KdV 
equation (where the dependent variable has values in $\mathbb{C}$).  
 
\diagramstyle[PostScript=dvips]
\newarrow{Dashto}{}{dash}{}{dash}>
\begin{diagram}[notextflow]
 \hat{\cal S}_1 & \rTo^{\widehat{BT}_{21}} & \hat{\cal S}_2 & \rTo^{\widehat{BT}_{32}} & \hat{\cal S}_3 \\
 \dTo^{\pi_1}   &                          & \dTo^{\pi_2}   &                          & \dTo^{\pi_3}   \\
 {\cal S}_1     & \rDashto                 & {\cal S}_2     & \rDashto                 & {\cal S}_3     \\ 
\end{diagram}
\begin{center}
{\bf Fig.~2:}  
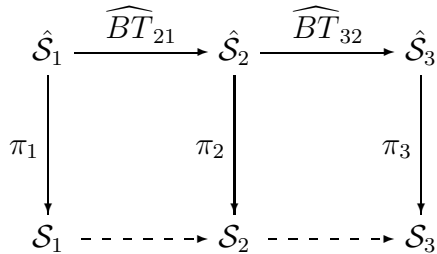
 BTs and reductions (the maps $\pi_i$).
\end{center}
\vskip.1cm

The method presented in this work is a constructive one, a receipe 
to determine BTs. We do not know, how exhaustive it is and the 
techniques developed here do not provide us with suitable tools 
to answer this question. Other techniques are available, of course, 
like those in the jet-bundle framework (see \cite{Roge+Shad82}, 
chap. 2, and \cite{KLV86}, for example). 
\vskip.1cm

In our examples, we have concentrated on equations in two (continuous 
or discrete) dimensions, with the exception of Hirota's difference equation 
which depends on three discrete variables. Of course, the method also 
applies to other higher-dimensional equations possessing a bicomplex 
formulation (cf \cite{DMH-bico,DMH-bidiff}). More examples of this 
kind will be studied elsewhere.

\end{document}